\RequirePackage[hyphens]{url}
\documentclass[twocolumn]{aastex631}

\PassOptionsToPackage{hyphens}{url}\usepackage{hyperref}
\definecolor{mylinkcolor}{RGB}{16, 37, 110}
\hypersetup{linkcolor=mylinkcolor,citecolor=mylinkcolor,urlcolor=mylinkcolor}

\shorttitle{HD~142527~B MagAO}
\shortauthors{Balmer et al.}

\received{September 2021}
\revised{May 2022}
\accepted{May 25th, 2022}
\submitjournal{The Astronomical Journal}

\graphicspath{{./}{figs/}}

\usepackage[nolist]{acronym}
\newacro{magao}[MagAO]{Magellan Adaptive Optics}
\newacro{visao}[VisAO]{Visible Light Adaptive Optics}
\newacro{ssdi}[S-SDI]{Simultaneous-Spectral Differential Imaging}
\newacro{sdi}[SDI]{Spectral Differential Imaging}
\newacro{adi}[ADI]{Angular Differential Imaging}
\newacro{rdi}[RDI]{Reference Differential Imaging}

\begin{document}

\title{Improved Orbital Constraints and H-alpha Photometric Monitoring of the Directly Imaged Protoplanet Analog HD~142527~B}

\author[0000-0001-6396-8439]{William O. Balmer}
\correspondingauthor{William O. Balmer}
\email{wbalmer1@jhu.edu}
\affiliation{Department of Physics \& Astronomy, Johns Hopkins University, 3400 N. Charles Street, Baltimore, MD 21218, USA}
\affiliation{Space Telescope Science Institute, 3700 San Martin Drive, Baltimore MD 21218, USA}

\author[0000-0002-7821-0695]{Katherine B. Follette}
\affiliation{Department of Physics \& Astronomy, Amherst College, 25 East Drive, Amherst, MA 01002, USA}

\author[0000-0002-2167-8246]{Laird M. Close}
\affiliation{Steward Observatory, University of Arizona, Tucson, 933 N Cherry Ave, Tucson, AZ 85721, USA}

\author{Jared R. Males}
\affiliation{Steward Observatory, University of Arizona, Tucson, 933 N Cherry Ave, Tucson, AZ 85721, USA}

\author[0000-0002-4918-0247]{Robert J. De Rosa}
\affiliation{European Southern Observatory, Alonso de C\'{o}rdova 3107, Vitacura, Santiago, Chile}

\author[0000-0002-4489-3168]{J\'ea I. Adams Redai}
\affiliation{Center for Astrophysics, Harvard \& Smithsonian, 60 Garden Street, Cambridge, MA 02138, USA}

\author{Alex Watson}
\affiliation{Department of Physics \& Astronomy, Amherst College, 25 East Drive, Amherst, MA 01002, USA}

\author[0000-0001-6654-7859]{Alycia J. Weinberger}
\affiliation{Earth \& Planets Laboratory, Carnegie Institution for Science, \\ 5241 Broad Branch Rd NW, Washington, DC 20015, USA}

\author[0000-0002-1384-0063]{Katie M. Morzinski}
\affiliation{Steward Observatory, University of Arizona, Tucson, 933 N Cherry Ave, Tucson, AZ 85721, USA}

\author[0000-0001-7525-7423]{Julio Morales}
\affiliation{Department of Astronomy, University of Massachusetts, Amherst, MA, 01003, USA}

\author[0000-0002-4479-8291]{Kimberly Ward-Duong}
\affiliation{Department of Astronomy, Smith College, Northampton MA 01063 USA}

\author[0000-0003-3818-408X]{Laurent Pueyo}
\affiliation{Department of Physics \& Astronomy, Johns Hopkins University, 3400 N. Charles Street, Baltimore, MD 21218, USA}
\affiliation{Space Telescope Science Institute, 3700 San Martin Drive, Baltimore MD 21218, USA}

\begin{abstract}
Companions embedded in the cavities of transitional circumstellar disks have been observed to exhibit excess luminosity at H$\alpha$, an indication that they are actively accreting. We report 5 years (2013-2018) of monitoring of the position and H$\alpha$ excess luminosity of the embedded, accreting low-mass stellar companion HD~142527~B from the MagAO/VisAO instrument. We use \texttt{pyklip}, a python implementation of the Karhounen-Loeve Image Processing algorithm, to detect the companion. Using \texttt{pyklip} forward modeling, we constrain the relative astrometry to $1-2~\mathrm{mas}$ precision and achieve sufficient photometric precision ($\pm0.2~\mathrm{mag}, 3\%$ error) to detect changes in the H$\alpha$ contrast of the companion over time. In order to accurately determine the relative astrometry of the companion, we conduct an astrometric calibration of the MagAO/VisAO camera against 20 years of Keck/NIRC2 images of the Trapezium cluster. We demonstrate agreement of our VisAO astrometry with other published positions for HD~142527~B, and use \texttt{orbitize!} to generate a posterior distribution of orbits fit to the relative astrometry of HD~142527~B. Our data suggest that the companion is close to periastron passage, on an orbit significantly misinclined with respect to both the wide circumbinary disk and the recently observed inner disk encircling HD~142527~A. We translate observed H-alpha contrasts for HD~142527~B into mass accretion rate estimates on the order of $4-9\times10^{-10}~\mathrm{M_\odot}\mathrm{yr}^{-1}$. Photometric variation in the H-alpha excess of the companion suggests that the accretion rate onto the companion is variable. This work represents a significant step towards observing accretion-driven variability onto protoplanets, such as PDS~70~bc.
\end{abstract}


\NewPageAfterKeywords


\section{Introduction} \label{sec:intro}

\par As the circumstellar environment surrounding young pre-main-sequence stars evolves, forming planets and binary companions disrupt and shape the circumstellar disk \citep[e.g.][]{Williams2011, Dong2017}. Embedded giant planets (protoplanets) and binary companions are expected to play dramatic roles in the formation of substructures such as cavities, rings, and spiral features within disks \citep[e.g.][]{Dodson-Robinson2011, Bae2018}. In the past two decades, improvements in high-contrast imaging instrumentation and post-processing techniques have revealed these morphologically complex disks in striking detail \citep[e.g.][]{Tamura2016,Avenhaus2018,Esposito2020,Garufi2020}. Of particular interest are so-called ``transition disks" \citep{Dodson-Robinson2011, Espaillat2014}, which host a wide central cavity depleted of dust. Gas has been observed to flow through these cavities \citep{Casassus2013}, fueling accretion onto the central star as well as onto the planetary \citep[e.g.][]{Sallum2015, Wagner2018, Haffert2019, Zhou2021} and stellar \citep{Close2014} mass companions that may be responsible for clearing them. This accretion is likely mediated by a circumsecondary (or circumplanetary) disk, observational evidence of which is accumulating at NIR \citep[e.g.][]{Lacour2016} and sub-mm \citep[e.g.][]{Benisty2021} wavelengths.
\par Cleared central cavities and ongoing accretion make transition disks prime targets for visible light adaptive optics imaging searches for forming protoplanets. Accreting companions are expected to exhibit substantial H$\alpha$ excess emission during accretion \citep{Mordasini2017, Szulagyi2020, Aoyama2021}. This enables their detection at lower contrasts than non-accreting objects. Taking images through H$\alpha$ and nearby continuum filters simultaneously and subtracting them allows for measurement of the H$\alpha$ luminosity of the accreting object through \ac{ssdi}. So far, few\footnote{Aside from HD~142527~B, there are a total of four candidate accreting objects currently known to orbit \textit{within} transitional disk cavities. They include the (disputed) candidate LkCa 15b: $\mathrm{M_p}<5-10\mathrm{M_J}$, $\mathrm{M_p\dot{M}}\sim10^{-6}\mathrm{M_J}^2\mathrm{yr}^{-1}$ \citep{Sallum2015, Currie2019}, confirmed planet PDS~70~b: $\mathrm{M_p}\sim4-10\mathrm{M_J}$, $\mathrm{\dot{M}}\sim5\times10^{-7}\mathrm{M_J}\mathrm{yr}^{-1}$ \citep{Keppler2018, Wagner2018, Haffert2019, Hashimoto2020, Wang2021}, confirmed planet PDS~70~c: $\mathrm{M_p}\sim4-12\mathrm{M_J}$, $\mathrm{\dot{M}}\sim1\times10^{-8}\mathrm{M_J}\mathrm{yr}^{-1}$ \citep{Haffert2019, Wang2021}, and candidate AB~Aur~b: $\mathrm{M_p}\sim9-12\mathrm{M_J}$ \citep{Currie2022}.} 
embedded, accreting companions have been directly imaged at H$\alpha$, with recent surveys returning no new detections \citep{Cugno2019, Zurlo2020}. This may be a low Strehl selection effect \citep{Close2020} or a reflection of differences in accretion physics at planetary masses \citep{Aoyama2021, Marleau2022}.

\section{HD 142527} \label{sec:litrev}
\par HD 142527 is a well-studied transitional disk system at a distance of $159.3\pm0.7\mathrm{pc}$ \citep{GaiaCollaboration2021}. The central star, HD~142527~A, is a young ($5.0\pm1.5~\mathrm{Myr}$; \citealp{Mendigutia2014}) F6III-V type Herbig Ae/Be star with an estimated mass of $\mathrm{M_A}=2.0\pm0.3~\mathrm{M_\odot}$ \citep{Mendigutia2014}. It exhibits accretion on the order of $10^{-7}~\mathrm{M_\odot}\mathrm{yr}^{-1}$ that is variable by at least a factor of 7 over $\sim5$ years \citep{Mendigutia2014}. The system has a low extinction, \citep[$A_v = 0.0^{+0.05}_{-0.00}$, ][]{Fairlamb2015}. The outer disk surrounding HD~142527~A is moderately inclined  \citep[$\mathrm{i_{disk}}=28.0\pm0.5^\circ$][]{Perez2015, Boehler2017}, hosts spiral arms in both scattered light \citep[e.g.][]{Fukagawa2006, Avenhaus2014, Hunziker2021} and sub-mm gas emission \citep[e.g.][]{Garg2021, Boehler2021}, and exhibits an asymmetrical `horseshoe' of sub-mm thermal emission \citep[e.g.][]{Ohashi2008, Boehler2017}. The central cavity of the disk is heavily depleted of dust \citep{Avenhaus2017} and extends to $\sim140$au \citep{Avenhaus2014}. Gas emission within the cavity exhibits a complex morphology, with a possible warp in the innermost region \citep{Casassus2013, Casassus2015, Perez2015} and non-Keplerian motion throughout \citep{Garg2021}.

\par The inner thermally-emitting disk component was first inferred by SED modeling of NIR excess and unresolved sub-millimeter observations \citep{Verhoeff2011, Boehler2017}.  Narrow ``shadow" features have been observed along the outer disk cavity wall, suggesting that this inner disk may be inclined with respect to the outer disk \citep{Marino2015}. \citet{Bohn2021} provided the first resolved measurements of the inner disk using VLTI/GRAVITY observations, which they used to investigate the mutual alignments of the inner and outer disk components. They found that the K-band complex visibilities of their data were best fit by an inner disk model with an inclination of  $\mathrm{i_{d, inn}} = 23.76\pm3.18^\circ$ and a longitude of ascending node of $\Omega\mathrm{_{d, inn}}=15.44\pm7.44^\circ$. Refitting ALMA CO data to derive outer disk geometry, they found an outer disk inclination of $\mathrm{i_{d, out}} = 38.21\pm1.38^\circ$ and a longitude of ascending node of $\Omega\mathrm{_{d, out}}=162.72.44\pm1.38^\circ$. From these measurements they inferred that the inner and outer disks of HD142527 are statistically significantly misaligned by $59^\circ$. This difference in inclination is consistent with the inner disk generating the shadows observed in scattered light and is suggestive of dynamical disruption, namely a companion on an inclined orbit \citep{Facchini2018}.

\newacro{sam}[SAM]{Sparse Aperture Masking}
\newacro{cadi}[cADI]{classical Angular Differential Imaging}

\par In 2012, a companion candidate to HD~142527~A was detected with the VLT/NACO instrument \citep{Lenzen2003, Rousset2003} using \ac{sam} at $88~\mathrm{mas}$ separation, with an estimated mass of $0.1-0.4~\mathrm{M_\odot}$ \citep{Biller2012}. Subsequently, the \ac{magao} team used the \ac{visao} instrument to confirm the companion by directly imaging HD~142527~B in H$\alpha$ ($\Delta\mathrm{mag}=6.33\pm0.20~\mathrm{mag}$) and in 643nm continuum ($\Delta\mathrm{mag}=7.50\pm0.25~\mathrm{mag}$) with a combination of \ac{cadi} and \ac{ssdi} \citep[referred to hereafter as ASDI, ][]{Close2014}. The presence of H$\alpha$ excess in the \ac{visao} detection indicated that the companion was actively accreting at a rate of $\sim\mathrm{\dot{M}}=5.9\times10^{-10}~\mathrm{M_\odot}\mathrm{yr}^{-1}$ \citep{Close2014}.

\newacro{pca}[PCA]{Principle Component Analysis}
\newacro{pdi}[PDI]{Polarized Differential Imaging}

\par The companion was also imaged using the Gemini Planet Imager \citep[GPI, ][]{Macintosh2014} in total intensity at Y-band with \ac{adi}/\ac{pca} and polarized intensity light with \ac{pdi} and \ac{adi}/\ac{pca} \citep{Rodigas2014}. Interestingly, the polarized light detection was marginally spatially inconsistent with the total intensity source by $\sim20~\mathrm{mas}$ at $2\sigma$ confidence in their reduction, which they interpret either as scattered light from the disk around the companion, or a clump of dust separated from the companion\footnote{It should be noted that \citet{Avenhaus2017} do not detect a compact polarized source in their SPHERE/ZIMPOL images of the circumprimary environment.}.
\par The companion was subsequently characterized using VLT/NACO and Gemini/GPI \ac{sam}; it was confirmed to exhibit infrared excess indicative of a 1700K circumsecondary environment and was found to be significantly younger ($1.0\pm1\mathrm{Myr}$) than A \citep{Lacour2016}. \citet{Lacour2016} fit a mass of $\mathrm{M_B}=0.13\pm0.03~\mathrm{M_\odot}$ and a temperature of $\mathrm{T_B}=3000\pm100\mathrm{K}$ to the SED of the object. Using the SPHERE/SINFONI instrument \citep{Eisenhauer2003}, \citet{Christiaens2018} find $\mathrm{T_B}=3500\pm100$~K, a spectral type of M2.5, and therefore an age of $0.75\pm0.25\mathrm{Myr}$ and mass of $\mathrm{M_B}=0.35~\mathrm{M_\odot}$ with a recovered spectrum that is significantly brighter than that found in \citet{Lacour2016}. They do not investigate this discrepancy\footnote{This may be due to a distance scaling error; see discussion in \citet{Greenbaum2019}.}. 
\par The SPHERE IFS and IRDIS instruments were used to examine the companion via NIR direct imaging and \ac{sam} \citep{Claudi2019}. Their results suggest best fit temperatures in the range of $2600-2800\mathrm{K}$, closer to those fit in \citet{Lacour2016}, and a spectral type M5-6. They also demonstrate variability in the differential flux of B with respect to the brightness of A on the order of half a magnitude between $1-1.6\mu$m. While this could be due to the variability of the primary star, \citet{Claudi2019} quantified the variability of the primary and found it was insufficient to explain the variability in the continuum flux of HD~142527~B. They place dynamical constraints on the mass of HD~142527~B, finding $\mathrm{M_B}=0.26^{+0.16}_{-0.14}~\mathrm{M_\odot}$.
\par \citet{Cugno2019} observed HD~142527~B using SPHERE/ZIMPOL in H$\alpha$ with \ac{adi}+\ac{ssdi} in a similar fashion to \citet{Close2014}. They recovered the companion in three filters -- a broad H$\alpha$ filter, a narrow H$\alpha$ filter, and a continuum filter. They confirmed H$\alpha$ excess emission from the companion and estimated an accretion rate of $\sim\mathrm{\dot{M}}=1-2\times10^{-10}~\mathrm{M_\odot}\mathrm{yr}^{-1}$, marginally lower than that derived by \citet{Close2014}.
\par Further study of HD~142527~B informs a number of open questions in the fields of star and planet formation. The extreme mass ratio ($\sim20:1$) between the primary and secondary in this system, the fact that HD~142527~B is orbiting within a transition disk cavity, and its ongoing accretion make the system a higher mass analog to protoplanetary systems such as PDS~70~b\&c, allowing for the refinement of techniques used to image these systems in visible light. 
\par The HD 1425257 system itself is also an important probe of the processes of planet formation in binary systems. Improved orbital constraints can place limits on the mutual inclinations between the HD~142527~AB binary orbit and the inner and outer disk segments \citep{Czekala2019}. Recent VLTI/GRAVITY observations of the inner disk are especially important to consider, as it is likely that HD~142527~B is responsible for the inclination of the misinclined inner disk. Orbit fits to the HD 142527 binary allow mutual inclinations of all three components of the system (the inner disk, binary, and outer disk) to be determined and compared to hydrodynamical models.
\par The nature of the HD~142527~B companion itself is also broadly informative of star and planet formation processes, as its young age relative to HD~142527~A might indicate that it formed from the disk via disk instability. It could also be that HD~142527~B formed at the lower end of the IMF, was dynamically captured, and is disrupting planet formation around HD~142527~A. In either scenario, its motion necessarily drives the dynamical evolution of the disk \citep[e.g.][]{Aly2021}.
\par Observations of the companion to date have only covered a $60-70^\circ$ orbital arc. The nature of HD~142527~B's orbit and its relationship to both the wide observed cavity in the circumbinary disk and the inner circumprimary disk is still an area of ongoing study that can be improved by further constraining the orbital elements, particularly the eccentricity and inclination of the binary orbit and the mutual inclinations of the various components. Until the orbit is very well characterized, it cannot be certain whether the companion can be held solely responsible for the massive cavity.
\par Additionally, relatively little is known about the density and dynamics of gas flow through transitional disk cavities or the reprocessing of material in circumsecondary accretion disks. Accretion rates and epoch-to-epoch variations in accretion rate can inform accretion processes for companions in these actively planet-forming systems.
\par In this work, we build upon previous detections of H$\alpha$ excess emission from HD~142527~B. We monitor the companion's astrometric motion in visible light and provide the most complete orbital solution to date, enabling a more precise comparison between the mutual inclinations of system components. We leverage additional epochs of observation at H$\alpha$ to argue that the accretion onto the companion is likely variable on (at least) yearly timescales.

\section{Observations \& Data Reduction} \label{sec:observations}
\subsection{VisAO observations}
\par We observed HD 142527 with the Magellan Adaptive Optics \citep{Close2013, Morzinski2016} \ac{visao} \citep{Males2014} instrument in H$\alpha$ SDI mode \citep{Close2014b} on 7 nights between 2013 and 2018. Table \ref{tab:visaodata} records general information about the \ac{visao} observations of HD 142527. During observations, the telescope rotator was turned off, resulting in rotation of the FOV, which enables \ac{adi}. More field rotation allows for more reference images to be used in constructing the PSF model, improving its quality; this is especially important for tightly-separated companions like HD~142527~B. The total field rotation for each observation is noted in Table \ref{tab:visaodata}. HD~142527~A was dithered across the CCD throughout the observing night to mitigate the effects of near-focus dust spots in the images. MagAO uses a Natural Guide Star (NGS) to conduct wavefront corrections, and in all datasets the NGS used was the on-axis science target, HD~142527~A. The SDI observing mode splits the incoming light using a Wollaston beamsplitter, and results in simultaneous continuum ($\lambda_c = 656\mathrm{nm}$, $\Delta\lambda=6.32\mathrm{nm}$) and H$\alpha$ ($\lambda_c = 643\mathrm{nm}$, $\Delta\lambda=6.20\mathrm{nm}$) images ($512\times1024$ pixels in size) on the top and bottom of the CCD, respectively.

\par We conducted a revised calibration of the absolute astrometric solution for the VisAO instrument, which is documented in detail in Appendix \ref{sec:absastr}. We determined an updated platescale and North Angle offset by tying observations of the $\theta^1$ Ori B multiple system over 6 years to observations of the same system taken over 20 years with the precisely astrometrically calibrated Keck/NIRC2 instrument \citep{Yelda2010, Service2016}. The updated \ac{visao} platescale is $7.95\pm0.010 ~\mathrm{mas}$ $\mathrm{pix}^{-1}$ and the updated North Angle offset is $0.497\pm0.192^\circ$ counterclockwise. This new solution does not necessitate a revision of previously published results, as both values agree with previous calibrations within errorbars \citep{Close2013, Males2014}. However, the updated calibration improves the accuracy of VisAO astrometry, and validates the stability of the North Angle offset between instrument mountings, which occur every semester.

\begin{deluxetable*}{ccccccccc}
\tablecaption{VisAO observations of HD 142527\label{tab:visaodata}}
\tablewidth{0pt}
\tablehead{
\colhead{Date} & \colhead{$\mathrm{N_{ims}}$} & \colhead{Exposure} & \colhead{Total integration} & \colhead{Total rotation} & \colhead{FWHM} & \colhead{Saturation radius} & \colhead{Avg. Seeing} & \colhead{Stellar H$\alpha$/Cont.}  \\
\colhead{(YY/MM/DD)} & \colhead{} & \colhead{(s)} & \colhead{(m)} & \colhead{($^\circ$)} & \colhead{(mas)} & \colhead{(pix)} & \colhead{('')} & \colhead{} \\
}
\startdata
2013-04-11 & 1961 & 2.27     & 74.2 & 65.3  & 25.0 & \nodata& \nodata & $0.83\pm0.03$ \\
2014-04-08 & 1758 & 2.27     & 66.5 & 101.7 & 38.9 & \nodata& \nodata & $1.11\pm0.10$ \\
2015-05-15 & 2387 & 2.27     & 90.3 & 117.4 & 37.8 & \nodata& 0.55    & $1.19\pm0.23$ \\
2015-05-16 & 1143 & 2.27& 43.2 & 34.8  & 33.8 & 2 & 0.80    & $1.17\pm0.07$ \\
2015-05-18 & 159  & 30   & 79.5 & 76.8  & 30.7 & 9 & 0.66    & $1.16\pm0.08$ \\
2017-02-02 & 242  & 12       & 48.4 & 16.1  & 43.7 & 2 & 0.72    & $1.13\pm0.02$ \\
2018-04-27 & 580  & 5    & 48.3 & 49.2  & 28.5 & 3 & \nodata & $1.31\pm0.10$
\enddata
\tablecomments{The average seeing was determined by measurements taken from the DIMM, Magellan Baade, or both (in which case the two are averaged). For datasets where no external seeing information exists, the column is left blank. The stellar H$\alpha/$Cont. ratios are calculated by performing aperture photometry on every image in both wavelengths for each dataset.  We extract photometry from the central star in the case of unsaturated datasets and from the instrumental ghost in the case of saturated data.  We report the median ratio for each dataset in the table above. A ratio $<1$ indicates a relatively quiescent phase in the accretion onto the primary. 
}
\end{deluxetable*}

\vspace{-1cm}
\subsection{Preprocessing}
\par \ac{visao} data were reduced and preprocessed with the GAPlanetS data reduction pipleline described in detail in \citet[]{Follette2017} and Follette et al. (2022, in prep.). In short, images were dark subtracted, flat fielded (except for the 2013-04-11 dataset, for which no flatfields exist), split between H$\alpha$ and continuum channels, and registered using a Fourier cross-correlation algorithm, which yields errors on the position of the central star of $\sim0.1\mathrm{pix}$ (Follette et al. 2022, in prep.). 

\par In half of our observations (3 of 7, see Table \ref{tab:visaodata}), the central star was allowed to saturate in order to improve SNRs in the search for additional companions within the system. A stable instrumental ghost appears in VisAO images to the right of the natural guide star (NGS). We investigated the stability of the ghost as a photometric and astrometric calibrator for saturated data, and found it to be suitable for both purposes (see Appendix \ref{sec:ghost}).

\par Following alignment, images with cosmic rays within $50\mathrm{pix}$ ($\approx0\farcs04$) of the central star were rejected by hand. Images were then cropped to a 451 pixel ($\sim$3$\farcs$5) square surrounding the central star. 
\par We then measured the ``H$\alpha$ scale factor" by conducting aperture photometry on the central star (or ghost, for saturated images) for each image in the sequence. We adopt the median of the ratios between the H$\alpha$ flux and the continuum flux for each image in the sequence as the best estimate of the scale factor ($F_{*,H\alpha}/F_{*,Cont}$), and the standard deviation as the uncertainty, and these values are recorded in Table \ref{tab:visaodata}. This scale factor is an estimate of the H$\alpha$ excess of the primary star, and allows us to quantify the accretion variability of the host star itself and to correct for the effect of stellar accretion on measured contrasts for the companion (see Section \ref{sec:analysis}).

\subsection{PSF Subtraction} \label{sec:psfsub}

\begin{figure*}
\centering
\includegraphics[width=\textwidth]{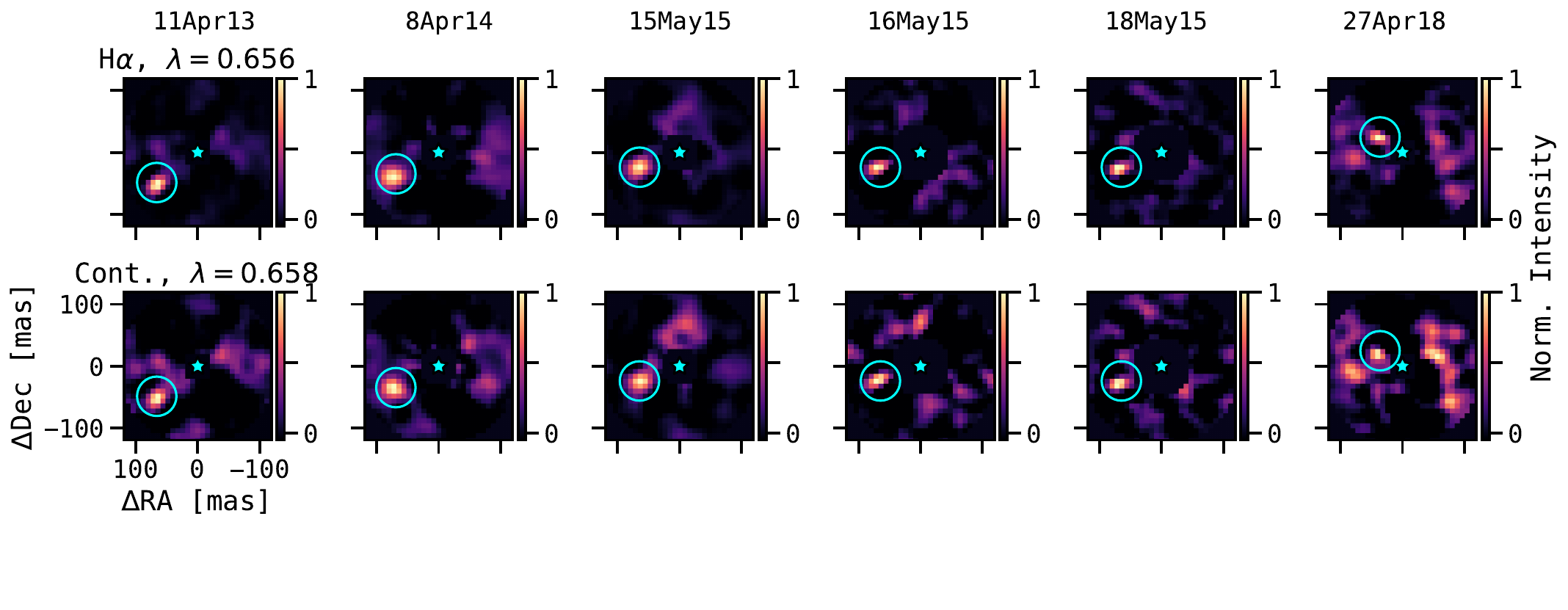}
\caption{Gallery of post-KLIP images showing the detections of HD~142527~B in H$\alpha$ (top) and continuum (bottom). The colorbar normalized to the peak pixel value of the companion in each image. The cyan circle indicates the nearest in time position of HD~142527~B reported in previous literature \citep{Lacour2016, Claudi2019}. The cyan star indicates the position of HD~142527~A. The innermost pixels have been masked to $r\sim1\times$FWHM for each dataset.}
\label{fig:SDIgallery}
\end{figure*}

\par We conducted PSF subtraction using the python implementation of Karhunen-Loeve Image Processing, \texttt{pyklip}\footnote{\url{https://pyklip.readthedocs.io}} \citep{Wang2015}. The \texttt{pyklip} input parameters \texttt{movement}, \texttt{annuli}, and \texttt{numbasis} were chosen based on the optimization techniques described in Adams Redai et al. (2022, in prep) as developed for the the Giant Accreting Protoplanets Survey (Follette et al. 2022, in prep). In brief, the \texttt{pyklip} \texttt{movement} parameter is an exclusion parameter that removes images from the reference library in which a companion at a given separation from the star would have shifted (rotated with the sky) by fewer than the specified number of pixels relative to the image for which the PSF model is being constructed. The \texttt{annuli} parameter describes the size of annular zones that \texttt{pyklip} treats separately, where the width of the annulus is $\Delta r=(\mathrm{OWA-IWA})/\texttt{annuli}$. Generically, the inner edge of each annulus is defined as $ r_{in} = \textrm{IWA} + \Delta r \times n$, but for HD~142527~B, the companion always lies within the inner annulus, so n=1 and $r_{in} = IWA$, while $r_{out}= IWA + \Delta r$.  Our images are non-coronagraphic, so do not have a hardware-determined inner working angle. However, we found that applying one in software generally improves our detections and we have adopted a fixed value of $1\times\mathrm{FWHM}$ here. The \texttt{numbasis} parameter controls the number of principle components, or KL modes, included in the constructed PSF. We chose to fix the maximum number of KL basis vectors used to construct the PSF at 100 and applied a pre-KLIP highpass filter of $1\times\mathrm{FWHM}$ in all analyses to match the broader GAPlanetS Survey strategy.
\par We refer the reader to Adams Redai et al. (2022, in prep) for additional information on the optimization of \texttt{pyklip} parameters for GAPlanetS data. To briefly summarize the optimization method used here, we:
\begin{enumerate}
    \item conduct a grid search in \texttt{pyKLIP}'s \texttt{movement}, \texttt{annuli}, and \texttt{numbasis} (KL mode) parameters and generate KLIP images for each combination
    \item compute six ``image quality metrics" for each \texttt{movement}, \texttt{annuli}, and \texttt{numbasis} combination. These metrics are: peak SNR of the companion(s), average SNR of the companion(s), the ``neighbor quality" of the previous two metrics (computed by smoothing their metric maps), false positive ($>$5$\sigma$) pixels, and contrast, normalized so that their best (highest for SNRs, lowest for contrast and false positive pixels) values are 1.
    \item Sum or average across a desired choice of metrics, companion(s), and KL modes to select ``optimal" values for each of the three KLIP parameters.
\end{enumerate}

For this work, the target of the optimization for most datasets were ``false planets" inserted into into the continuum images. In these datasets, we optimized on the sum of all 6 normalized image quality metrics averaged between 5 and 20 KL modes and across 4 to 8 false planets (as many as would fit with a radial spacing of 0.5$\times$FWHM and an azimuthal spacing of 85$^o$ between the IWA and control radius of each dataset) to select the optimal \texttt{movement} and \texttt{annuli} parameters. We then selected a single optimal \texttt{numbasis} value by maximizing the sum of the 6 normalized metrics for the optimal combination of \texttt{annuli} and \texttt{movement}. The optimal values for all three parameters are reported in Table \ref{tab:klipparams}.

Although the companion is recovered in all epochs using this parameter selection method, which optimizes for robust recovery of companions throughout the region between the IWA and control radius, it's not necessarily the best choice of parameters for the specific location of the companion. In the case of a robust known companion like HD~1425257~B, optimization can also be done on the location of the companion itself in H$\alpha$ images, and we utilize this method to achieve higher SNR recoveries of the companion in the May 16, 2015 and April 27, 2018 datasets. For these epochs, we optimize on the sum of the peak and average SNR metrics for the companion, and select the \texttt{annuli}, \texttt{movement}, and \texttt{numbasis} parameters that maximize this sum. Since each epoch of VisAO data for HD~142527~B has a corresponding near-in-time NIR detection, we optimized the KLIP parameters on the known NIR location of the companion. Contrast curves and limits on additional companions in the HD 142527 cavity appear in the upcoming GAPlanetS survey paper, Follette et al. (2022, in prep).


\begin{deluxetable*}{ccccc}

\tablecaption{Adopted \texttt{pyklip} starlight subtraction parameters\label{tab:klipparams}}
\tablewidth{0pt}
\tablehead{
\colhead{Date} & \colhead{Movement} & \colhead{Annuli} & \colhead{KL mode}  \\
}
\startdata
2013-04-11  & 1        & 14      & 20  \\
2014-04-08  & 1        & 25      & 100 \\
2015-05-15  & 2        & 20      & 20  \\
2015-05-16  & 12       & 1       & 12  \\
2015-05-18  & 7        & 2       & 5   \\
2018-04-27  & 1        & 25      & 20 
\enddata
\end{deluxetable*}

\vspace{-1cm}
\section{Results} \label{sec:results}
\par Using \texttt{pyklip}, we detect the companion in both filters in all epochs of observation except 2017-02-02, where the total field rotation was too small ($16.1^\circ$) to achieve the necessary contrast to detect the very tightly separated companion\footnote{At the expected separation in this epoch \citep[50mas,][]{Claudi2019}, the highest contrast observable within this dataset was $0.5\times10^{-2}$, whereas our \textit{brightest} detection of HD~142527~B across all epochs was at a contrast of $0.3\times10^{-2}$.}. As expected, the observations with the greatest total field rotation (2015-05-15) and best atmospheric quality (2013-04-11) yielded the highest SNR recoveries of the companion. Figure \ref{fig:SDIgallery} shows each recovery in H$\alpha$ and continuum. We quantify the quality of each recovery using the forward modeling capabilities of \texttt{pyKLIP}.

\subsection{Bayesian Forward Modeling} \label{sec:bka}
\newacro{bka}[BKA]{Bayesian Klip Astrometry}
\par To determine the astrometry of the imaged companion, we implement KLIP forward modeling \citep{Pueyo2016} and conduct Bayesian Klip Astrometry \citep[BKA,][]{Wang2016}. This technique involves the projection of a companion PSF estimate onto the basis vectors used to construct the primary KLIP PSF model to synthesize a forward-modeled PSF. This forward-model is fit to the post-KLIP data using an affine-invariant MCMC with \texttt{emcee} \citep{emcee}. This produces a posterior distribution of model fits that yields  robust uncertainties on astrometry and photometry. We take the median of the posterior distribution of a given parameter as its measured value and the 16th and 84th percentiles as uncertainties on that median. The strength of KLIP forward modeling when compared to negative planet injection grid searches is the speed and precision of model fitting and convergence, which enables the application of a Bayesian analysis framework as well as the capability to model correlated noise within the images using a gaussian process.
\par For unsaturated data, we find that using a 2D Gaussian PSF with the FWHM of the median stellar PSF produces excellent forward model fits (see discussion in Appendix \ref{sec:ghost}). 
\par For saturated datasets, we do not have a direct measure of the stellar FWHM. In Appendix \ref{sec:ghost}, we investigate the suitability of a range of forward models in extracting astrometry and photometry for the tight HD42527B companion in saturated datasets. In short, we find that the ghost is suitable as a photometric calibrator (see Appendix \ref{sec:ghost}), and determine that the ratio of the peak value of a Moffat fit to the ghost to the peak value of a Moffat fit to the unsaturated NGS PSF in unsaturated data is a stable quantity. We also find that the ghost is slightly out of focus, and that its FWHM is $\sim7\%$ wider than the stellar FWHM in unsaturated datasets. This result matches our expectations: as the ghost is produced by a reflection off of the MagAO 50/50 beamsplitter, it has a longer path than the 0th order source, and is therefore slightly out of focus. We assume that both the star-to-ghost brightness scaling and the FWHM discrepancy holds in the case where the NGS is saturated, and estimate the FWHM of our saturated datasets by taking $0.93\times\mathrm{FWHM_{ghost}}$. We find that the optimal forward model for saturated data is a Gaussian PSF whose FWHM was set to the $0.93\times\mathrm{FWHM_{ghost}}$ to match the expected FWHM of the image plane PSF. 
\par In order to further quantify the strength of our detections, we utilize the \texttt{PlanetEvidence} class \citep{Golomb2019} within \texttt{pyklip} to conduct a Bayesian model comparison and determine the SNR of HD~142527~B for a given detection. \texttt{PlanetEvidence} uses the nested sampling implementation \texttt{pyMultiNest} \citep{Buchner2014, Feroz2009} to compare two models: $\mathrm{H_0}$, where the image contains only speckle noise, and $\mathrm{H_1}$, where the image contains a source at the position of HD~142527~B. \texttt{PlanetEvidence} returns marginal distributions of the parameters for the source and null cases, and calculates the SNR of the detection within the fitting region and the evidence values for $\mathrm{H_0}$ and $\mathrm{H_1}$ ($\mathrm{Z_0}$ and $\mathrm{Z_1}$). The log-ratio of these evidence values, $\log{B_{10}}=\log{Z_1/Z_0}$, enables us to quantify the confidence with which one model can be favored over the other.\footnote{See \citet{Golomb2019} for definitions of the evidence values, discussion of log-ratios, and an example involving quantifying the detection of $\beta$ Pictoris b.} This framework provides a more robust estimate of the quality of the detection, better capturing asymmetric speckle noise that can dominate at very close separations than SNR computed within an annulus. 
The SNR values reported in Table \ref{tab:astrometry} are calculated from the residuals within the fitting region as described in \citet{Golomb2019}.
\par For example, values for $\log{B_{10}}>5$ are considered ``strong" evidence against the null hypothesis. We achieved strong evidence ($\log{B_{10}}>10$) in all cases, and unambiguous detections ($\log{B_{10}}>20$) in eleven out of tweleve datasets. While the evidence is not strong for the detection in the continuum for the 2018-04-27 epoch, the search for the companion in this epoch is not a ``blind” search. We can be reasonably certain that signal at this location is due to the real companion due to the near-in-time observations in the NIR \citep{Claudi2019}. Nevertheless, the continuum detection in 2018 should be approached with some skepticism.

\subsection{Forward Model Results}
\par Our Bayesian forward modeling returns posterior distributions of the x and y positions (in pixels) of the fit model, a flux scaling parameter $\alpha$, and a length scale $l$ of correlated noise within the image. We transform $\alpha$ into contrast by multiplying it by a constant $10^{-2}$ (a \texttt{pyklip} input term setting the initial contrast of the forward model) and by the peak counts of the forward model (which in our case is 1 as the PSF model is normalized before it is passed to \texttt{pyklip}).

\par We calculate the astrometry given in Table \ref{tab:bkaresults} by applying the updated platescale and north angle offset (described in detail in Appendix \ref{sec:absastr}) to the separation and position angle measurements of the companion from the posterior distributions of forward model fits. We propagate $0.1\mathrm{pix}$ uncertainties on the position of the host star from the image registration process, $1\sigma$ uncertainties on the position of the companion from the posterior distribution of forward model fits, and estimated uncertainties on our absolute astrometry (see Appendix \ref{sec:absastr} for details) into our calculations to obtain final uncertainty estimates on separation and PA measurements for the companion.

\begin{figure*}
    \plotone{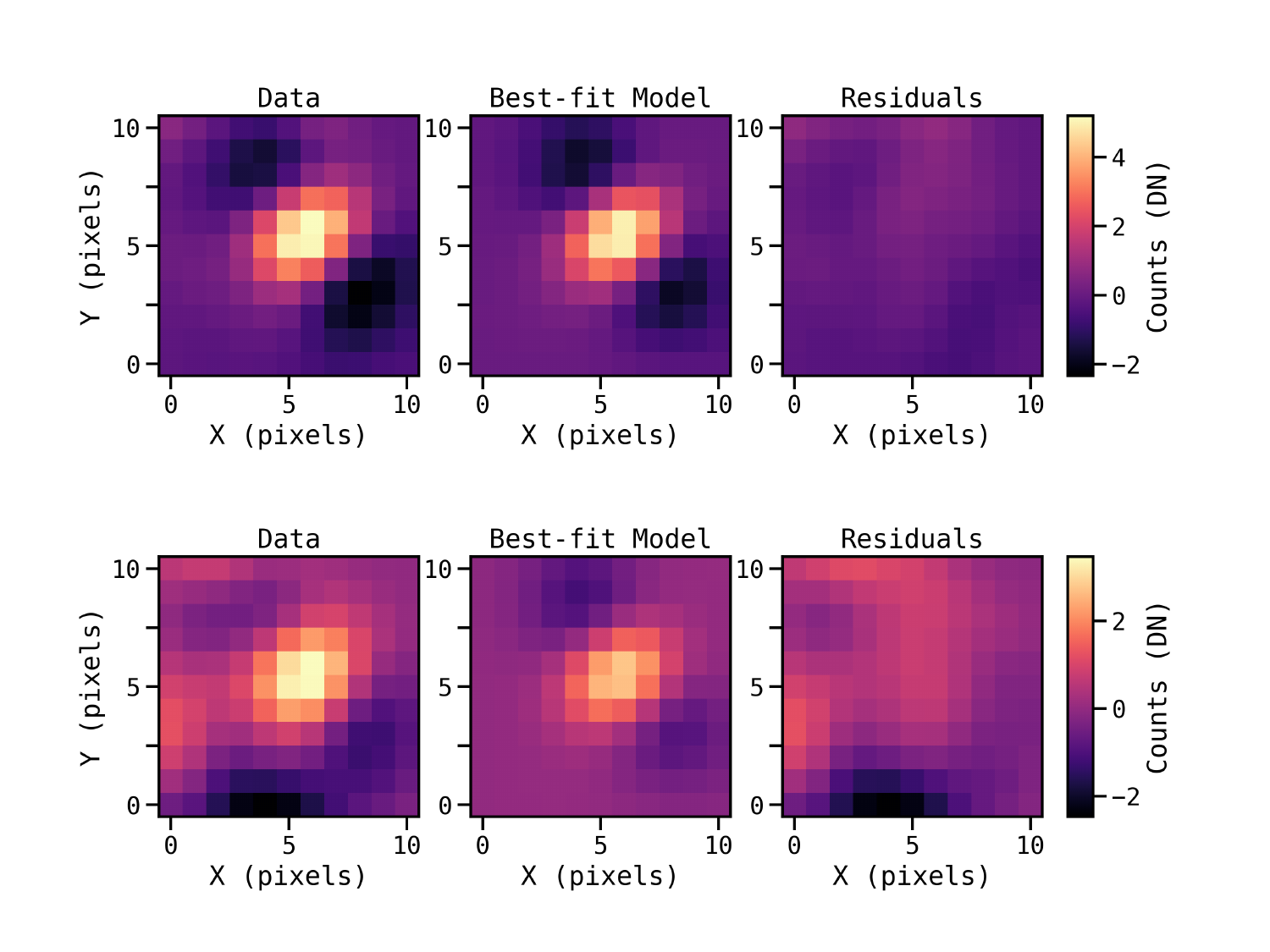}
    \caption{Representative \texttt{pyklip} forward model fits for HD~142527~B. Shown are the 2013-04-11 H$\alpha$ (top row) and continuum (bottom row) detections. Based on the \texttt{PlanetEvidence} analysis, the evidence ratios $\log{Z_1/Z_0}$ are 266.30 and 75.89 in H$\alpha$ and continuum, respectively. Both are considered extremely strong evidence in favor of the existence of the companion at this location.}
    \label{fig:exDataModelResid}
\end{figure*}

\begin{deluxetable*}{ccccccc}
\tablecaption{Results of Forward Model fitting and companion SNR \label{tab:bkaresults}}
\tablewidth{0pt}
\tablehead{
\colhead{Date} & \colhead{Separation} & \colhead{PA} & \colhead{$\Delta\mathrm{mag}$} & \colhead{$\log{Z_1/Z_0}$} & \colhead{SNR}\\
\colhead{(DD/MM/YY)} & \colhead{(mas)} & \colhead{($^\circ$)} & \colhead{(mag)} & \colhead{} & \colhead{}
}
\startdata
Continuum &               &               &             &        &       \\ \hline
2013-04-11 & $82.74\pm1.51$  & $128.07\pm0.68$ & $7.3\pm0.2$ & 75.89  & 5.24  \\
2014-04-08 & $78.25\pm3.39$  & $116.00\pm2.45$ & $7.2\pm0.3$ & 22.33  & 5.82  \\
2015-05-15 & $70.63\pm1.95$  & $112.28\pm1.38$ & $7.4\pm0.2$ & 22.77  & 7.85  \\
2015-05-16 & $70.59\pm4.36$  & $107.33\pm2.10$ & $7.4\pm0.3$ & 37.14  & 5.37  \\
2015-05-18 & $73.70\pm3.92$  & $110.97\pm2.24$ & $7.3\pm0.3$ & 77.43  & 5.98  \\
2018-04-27 & $42.75\pm3.22$  & $58.53 \pm2.95$ & $7.3\pm0.4$ & 13.55  & 2.64  \\ \hline
H$\alpha$  &                 &                 &               &        &       \\ \hline
2013-04-11 & $82.30\pm1.36$  & $127.91\pm0.61$ & $6.29\pm0.13$ & 266.30 & 10.53 \\
2014-04-08 & $77.57\pm1.86$  & $118.21\pm1.04$ & $6.59\pm0.17$ & 32.67  & 8.02 \\
2015-05-15 & $69.89\pm1.50$  & $109.68\pm0.98$ & $6.88\pm0.16$ & 51.26  & 12.03 \\
2015-05-16 & $72.64\pm2.29$  & $107.98\pm1.09$ & $6.73\pm0.19$ & 75.94  & 8.11 \\
2015-05-18 & $69.96\pm1.36$  & $110.02\pm0.76$ & $7.11\pm0.16$ & 131.84 & 7.90 \\
2018-04-27 & $44.29\pm2.57$  & $58.14 \pm2.20$ & $6.9\pm0.3$ & 24.31  & 3.47
\enddata

\end{deluxetable*}

\vspace{-0.85cm}
\par We determine the H$\alpha$ contrast of the companion relative to the primary star reported in Table \ref{tab:bkaresults} as follows. First, we take the median and standard deviation on the forward model fit to the contrast ($C_{B-A}$) as the derived value and uncertainty, respectively. For the continuum filter, this value can be converted to a contrast in magnitudes via a simple magnitude transformation, namely $\Delta\mathrm{mag}=-2.5\log{\mathrm{C_{B-A}}}$. However, at H$\alpha$, the star itself is actively accreting, influencing the H$\alpha$ contrast extracted from BKA. In order to measure the H$\alpha$ contrast of the companion \textit{with respect to the stellar continuum}, we multiply the BKA-derived contrast of the companion at H$\alpha$ ($F_{comp,H\alpha}/F_{*,H\alpha}$) by the stellar H$\alpha$ to continuum scale factor for the observations ($F_{*,H\alpha}/F_{*,Cont}$, determined as described in Section \ref{sec:observations} and given in Table \ref{tab:visaodata}). This leaves us with the contrast of the companion at H$\alpha$ \textit{relative to the stellar continuum} ($F_{comp,H\alpha}/F_{*,Cont}$) and allows us to compare the brightness of the companion at H$\alpha$ over time without being influenced by stellar H$\alpha$ variability. We propagate errors on the BKA-derived contrast and scale factor through the logarithmic transformation to magnitude.

Table \ref{tab:bkaresults} records the results of our forward model fits to the astrometry and photometry of continuum and H$\alpha$ images for each epoch. Figure \ref{fig:exDataModelResid} illustrates a representative model fit and residuals for the 2013-04-11 dataset, and galleries of forward model fits to other epochs can be found in Appendix \ref{sec:BKAfits}.

\section{Analysis \& Discussion} \label{sec:analysis}

\par
\begin{figure*}
\plotone{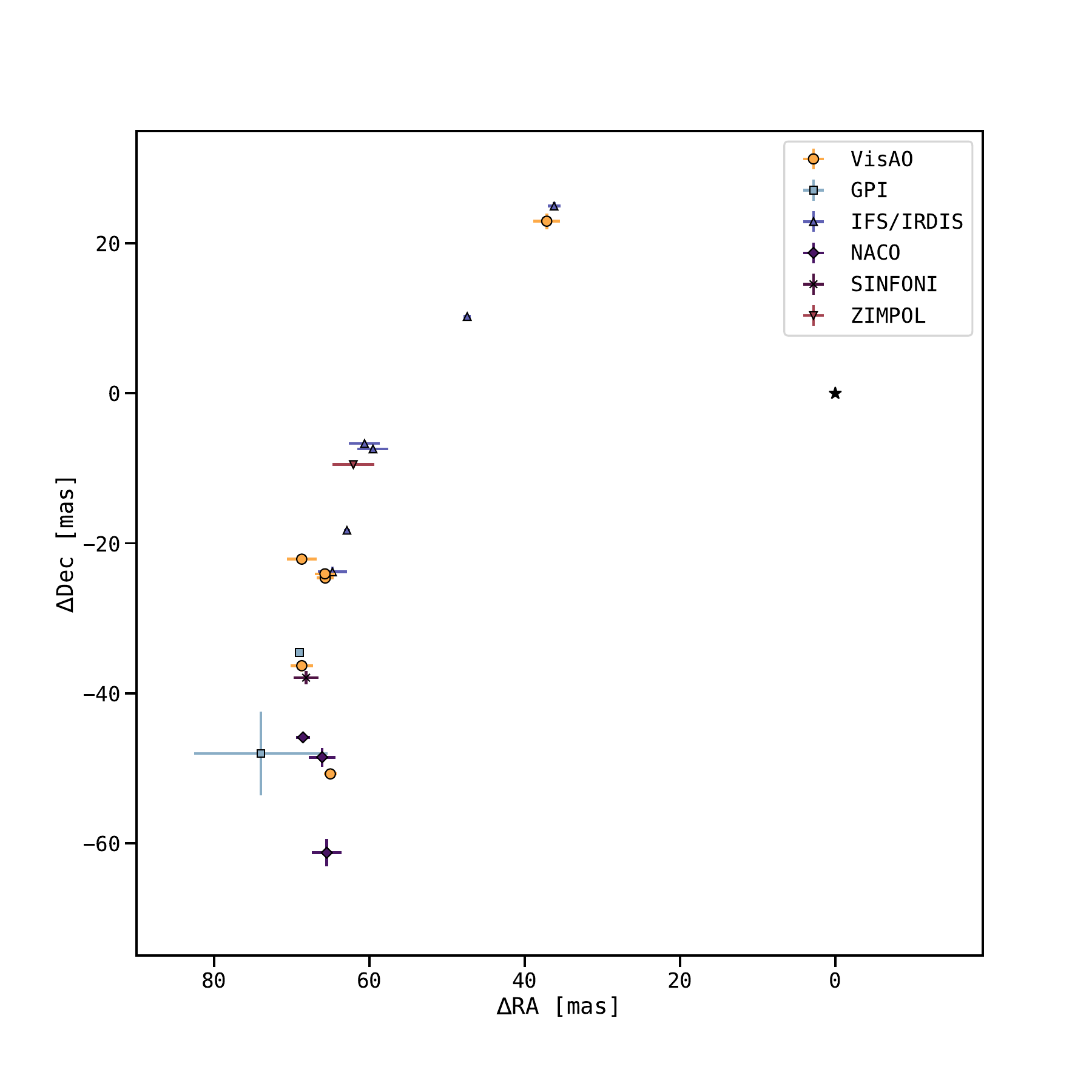}
\caption{Astrometric measurements of HD~142527~B relative to HD~142527~A (black star) from NACO, GPI, SPHERE, SINFONI, and VisAO (this work) between 2012 and 2018. The crosses represent the reported uncertainties on each measurement. The companion experiences significant orbital motion ($\Delta\theta\approx65^\circ$) and decreases substantially in separation ($\Delta\rho\approx30~\mathrm{mas}$) from 2016 to 2018. The VisAO values (yellow circles) agree well with astrometry from other instruments. Error bars on our measurements reflect the 1-2mas accuracy achievable using \texttt{pyklip} forward modeling.}
\label{fig:astrometry}
\end{figure*}

 We calculate the separation and PA of the companion on each night of observation as the weighted mean of the x and y positions in the H$\alpha$ and continuum filters, where the weights are the uncertainties in each filter (calculated as described in Section \ref{sec:results}). The final uncertainty on the H$\alpha$ and continuum averaged position is then the square root of the sum of the uncertainties on the individual measurements. Table \ref{tab:astrometry} records the final \ac{visao} astrometry together with previous astrometry from the literature, and these positions are plotted in Figure \ref{fig:astrometry}.

\begin{deluxetable*}{ccccc}
\tablecaption{HD~142527~B astrometry\label{tab:astrometry}}
\tablewidth{0pt}
\tablehead{
\colhead{Date} & \colhead{Separation} & \colhead{Position Angle} & \colhead{Instrument} & \colhead{Source} \\
\colhead{(DD/MM/YY)} & \colhead{(mas)} & \colhead{($^\circ$)} & \colhead{} & \colhead{}
}
\startdata
2012/3/11 & $89.70\pm2.60$  & $133.10\pm1.90$ & NACO       & B12       \\
2013/3/17 & $82.00\pm2.10$  & $126.30\pm1.60$ & NACO       & L16       \\
2013/4/11 & $82.50\pm1.01$  & $127.98\pm0.46$ & VisAO      & This work \\
2013/7/14 & $82.50\pm1.10$  & $123.80\pm1.20$ & NACO       & L16       \\
2014/4/8  & $77.73\pm1.63$  & $117.87\pm0.96$ & VisAO      & This work \\
2014/4/25 & $88.10\pm10.10$ & $123.00\pm9.20$ & GPI        & R14       \\
2014/5/10 & $78.00\pm1.80$  & $119.10\pm1.00$ & SINFONI    & Ch18      \\
2014/5/12 & $77.20\pm0.60$  & $116.60\pm0.50$ & GPI        & L16       \\
2015/5/13 & $69.00\pm2.00$  & $110.20\pm0.50$ & IFS/IRDIS  & Cl19      \\
2015/5/15 & $70.16\pm1.19$  & $110.56\pm0.80$ & VisAO      & This work \\
2015/5/16 & $72.19\pm2.02$  & $107.84\pm0.97$ & VisAO      & This work \\
2015/5/18 & $70.00\pm1.35$  & $110.12\pm0.72$ & VisAO      & This work \\
2015/7/3  & $65.50\pm0.40$  & $106.20\pm0.40$ & IFS/IRDIS  & Cl19      \\
2016/3/26 & $60.00\pm2.00$  & $97.10\pm0.50$  & IFS/IRDIS  & Cl19      \\
2016/3/31 & $62.80\pm2.70$  & $98.70\pm1.80$  & ZIMPOL     & Cu19      \\
2016/6/13 & $61.00\pm2.00$  & $96.30\pm0.50$  & IFS/IRDIS  & Cl19      \\
2017/5/16 & $48.50\pm0.50$  & $77.80\pm0.20$  & IFS/IRDIS  & Cl19      \\
2018/4/14 & $44.00\pm1.00$  & $55.40\pm0.40$  & IFS/IRDIS  & Cl19      \\
2018/4/27 & $43.69\pm2.01$  & $58.28\pm1.76 $ & VisAO      & This work
\enddata
\tablecomments{Sources for the listed astrometry in order of appearance are B12 = \citeauthor{Biller2012}, L16=\citeauthor{Lacour2016}, R14=\citeauthor{Rodigas2014}, Ch18=\citeauthor{Christiaens2018}, Cl19=\citeauthor{Claudi2019}, Cu19=\citeauthor{Cugno2019}.}
\end{deluxetable*}

\vspace{-1cm}
\subsection{Orbit Fitting}
\par By combining our derived astrometry with compiled results from the literature, we compute orbits fit to the motion of HD~142527~B with \texttt{orbitize!} \citep{Blunt2020}, an open-source python package that performs Bayesian orbit fitting for directly imaged companions\footnote{\url{https://orbitize.info}}. We use the parallel tempered \citep{ptemcee} affine-invariant \citep{emcee} MCMC sampler in \texttt{orbitize!}, in order to determine the posterior probabilities for 8 orbital parameters: semi-major axis ($a$), eccentricity ($e$), inclination angle ($i$), argument of periastron of the companion's orbit ($\omega$), longitude of ascending node ($\Omega$), epoch of periastron passage ($\tau$), system parallax ($\pi$), and total mass of the binary system ($\mathrm{M_{tot}}$).

\par We assume default \texttt{orbitize!} priors except on the total mass of the system and the system parallax. We set a Gaussian prior on the system parallax, $\mathcal{N}(\mu=6.356,\,\sigma=0.047)$, following the measured Gaia eDR3 parallax \citep{GaiaCollaboration2021}. We adopt a Gaussian prior on the total system mass ($\mathrm{M_{tot}}$) of $\mathcal{N}(\mu=2.3~\mathrm{M_\odot},\,\sigma=0.3~\mathrm{M_\odot})$ based on the results of \citet{Mendigutia2014} and \citet{Claudi2019}. We refer the reader to \citet{Blunt2020} for information about \texttt{orbitize!} default priors, as well as more detailed descriptions of the variables involved in the orbital solution. Posterior distributions are computed by first initializing an MCMC with 50 walkers and 20 temperatures for a 100,000 step burn-in phase per walker; the walker samples are discarded before a total of 5,000,000 additional samples are recorded to approximate the posterior.

\par
\begin{figure*}
    \plotone{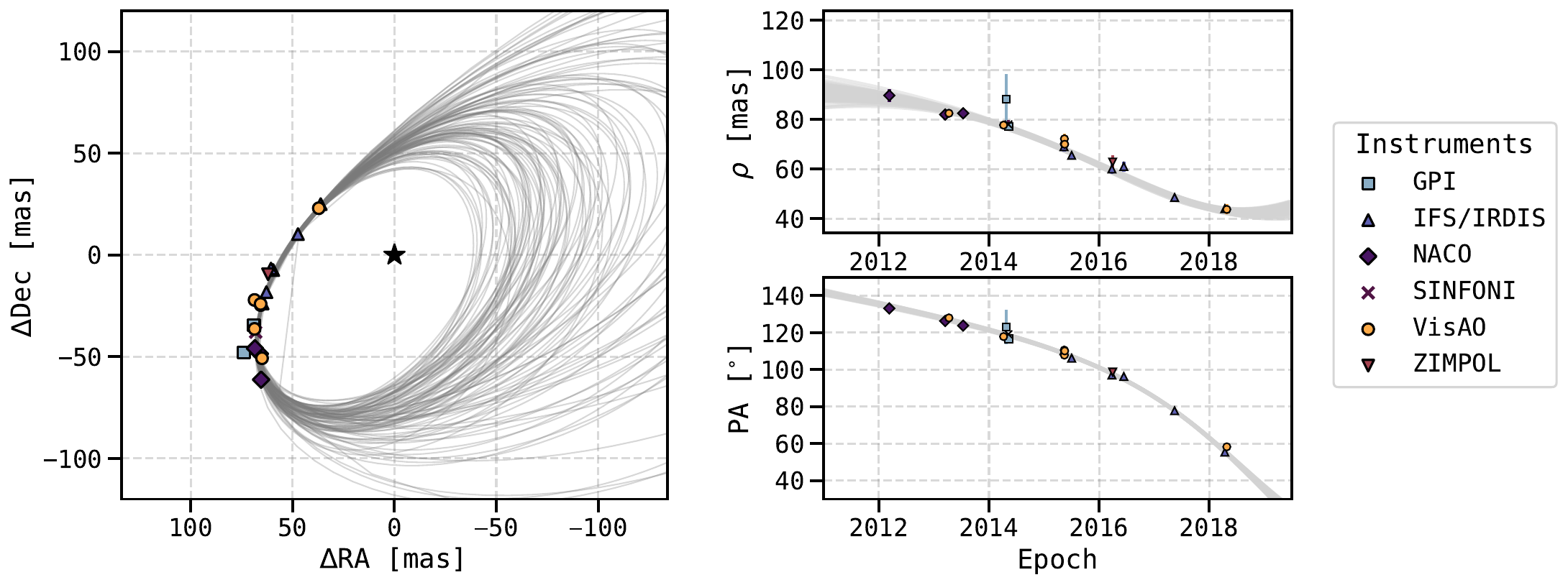}
    \caption{100 randomly drawn \texttt{orbitize!} orbits fit to the motion of HD~142527~B. Astrometry from Figure \ref{fig:astrometry} is overplotted. Left: the orbits projected in RA and Dec, relative to HD~142527~A (black star). Right: the orbits in separation/position angle versus time. The new VisAO astrometry does not add new coverage of the orbital arc, but the 2018 recovery adds more weight to the near-in-time SPHERE NRM astrometry.}
    \label{fig:orbitplot}
\end{figure*}

 Table \ref{tab:orbelems} records our estimates of the orbital elements for HD~142527~B. Figure \ref{fig:orbitplot} shows 100 randomly drawn orbits from our posterior distribution of orbit fits overplotted on the astrometry compiled in Table \ref{tab:astrometry}. The full corner plot visualizing the posterior distributions of all eight orbital parameters (Figure \ref{fig:142post}) can be found in Appendix \ref{sec:posteriors}. Our additional \ac{visao} astrometry provides marginally tighter constraints relative to previous orbital solutions. We note, however, that our MCMC approach is likely more robust under poorly-constrained posterior distributions than previous Least Squares Monte Carlo approaches \citep[See discussion in \S 2.3.2 of][]{Blunt2020}.

Our \texttt{orbitize!} orbit fitting yields well-converged unimodal distributions in all parameters except $\omega$ and $\Omega$, which show bimodal distributions with peaks spaced $180^\circ$ apart. This a known degeneracy in visual orbits with a lack of RV constraints\footnote{HD~142527~A is a pre-main-sequence F type star, which makes RV measurements particularly challenging. RV measurements for the system have been included in large pipeline surveys, such as the HARPS-RVBANK, but we choose not to consider these RVs for our orbital fitting, because they are likely dominated by stellar jitter which is outside of the scope of this paper to treat properly.}. To avoid confusion, we report values of the first ($0-180^{\circ}$) modes of $\omega$ and $\Omega$ in Table \ref{tab:orbelems}, noting that solutions with values 180$^{\circ}$ higher are equally likely.

\par Previous orbital solutions to the motion of HD~142527~B have have shown it to be both inclined \citep[$\mathrm{i}\sim125\pm5^\circ$][]{Lacour2016} and eccentric \citep[$\mathrm{e}>$0.2][]{Lacour2016,Claudi2019}. Our unimodal eccentricity distribution ($\mathrm{e}=0.24\pm0.15$) agrees with the first family of eccentricities fit by \citet[]{Claudi2019} ($\mathrm{e}\sim0.2-0.45)$; we do not reproduce their additional family of eccentricities ($\mathrm{e}\sim0.45-0.7$). This could be due to the increased weight our additional astrometry places on the position of the companion in 2018, or due to differences in the exploration of parameter space between the LSMC and MCMC frameworks.  We further constrain the system inclination to $\mathrm{i}=126\pm2^\circ$.

The longitude of ascending node ($\Omega$) and time of periastron passage of an orbit are more difficult to constrain with limited azimuthal coverage. \citet{Claudi2019} placed constraints of $124-135^\circ$ and 2020–2022 on these values. Our value of time of periastron passage, $2021.07^{+0.82}_{-0.72}$ agrees with their estimate within errorbars, but our refined $\Omega=142.38^{+5.51}_{-6.12}$ marginally disagrees; our solutions preferring a somewhat higher value. According to our orbital fits, periastron passage has likely already occurred for HD~142527~B.

\par HD~142527~B's orbit is interesting in the context of star and planet formation in that its eccentricity might indicate a stellar formation history. Directly imaged brown dwarf companions have been found to have a distribution of eccentricities that favor higher values when compared to directly imaged planets \citep{Bowler2020}, suggesting that they form independent of the system and are captured, rather than arising from within the circumplanetary disk. HD 142527 may therefore be more analogous to objects like GQ Lup B \citep{Wu2017, Stolker2021}, a captured, accreting Very Low Mass object that is dynamically truncating/disrupting its circumprimary environment, than to accreting protoplanets.

\begin{deluxetable}{cc}
\tablecaption{HD 142572B Orbital Elements\label{tab:orbelems}}
\tablewidth{0pt}
\tablehead{
\tablehead{\colhead{Element} & \colhead{Value}}
}
\startdata
a (au)                    & $14.71^{+8.18}_{-2.33}$                   \\
e                    & $0.28^{+0.22}_{-0.10}$                    \\
i $(^\circ)$                   & $126.27^{+2.13}_{-2.28}$                  \\
$\omega (^\circ)$                 & $86.02^{+52.31}_{-42.59}$  \\
$\Omega (^\circ)$                  & $142.38^{+5.51}_{-6.12}$ \\
Time of periastron passage (yr)                 & $2021.07^{+0.82}_{-0.72}$ \\
$\mathrm{M_{total}} (M_\odot)$                 & $2.49^{+0.27}_{-0.27}$ \\
Parallax (mas)                  & $6.36\pm0.05$ \\
$\theta_{in-\star} (^\circ)$ & $177.44^{+2.90}_{-2.80}, 110.26^{+1.82}_{-1.50}$  \\
$\theta_{\star-out} (^\circ)$ & $89.84^{+2.30}_{-1.65}, 158.82^{+2.76}_{-2.81}$  \\
\enddata
\end{deluxetable}

\vspace{-0.5cm}
\subsection{Mutual Inclinations}

\begin{figure*}
    \centering
    \includegraphics[width=\textwidth]{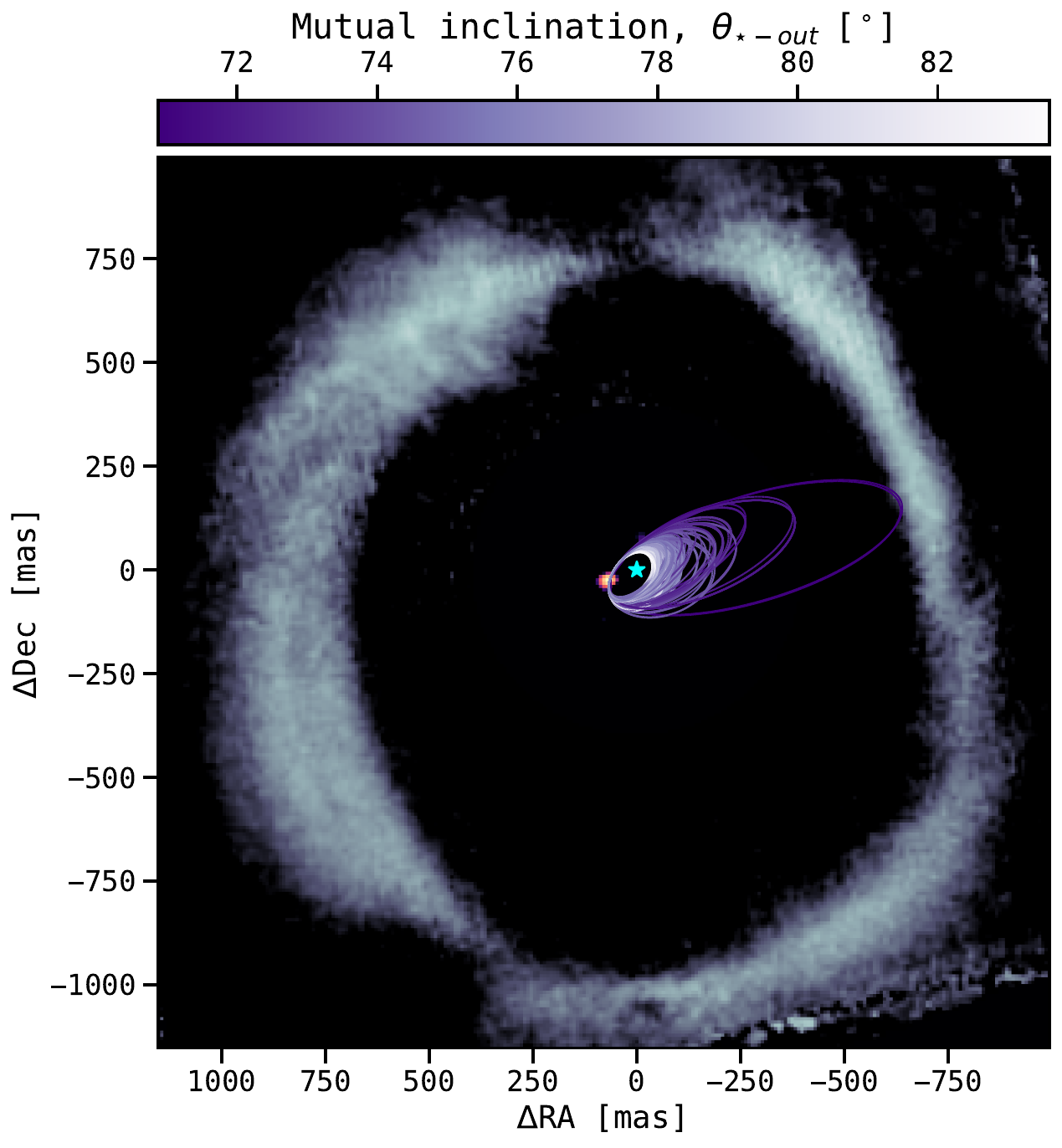}
    \caption{The orbit of HD~142527~B in context. The position of HD~142527~A is marked with a cyan star, and 250 randomly drawn orbits from fits to the astrometry of HD~142527~B are plotted with colors corresponding to their mutual inclination with the outer disk ($\theta_{\star-out}$). The outer image of the circumbinary disk is a Gemini/GPI scattered light polarized intensity H-band image reprocessed with an updated GPI pipeline and interpolated to VisAO's platescale. The inner image is the post-KLIP 2015/05/15 H$\alpha$ detection from VisAO (also shown in Figure \ref{fig:SDIgallery}). Both images are normalized to their respective maximum pixel values before combination.}
    \label{fig:disk_orb_det}
\end{figure*}

\par In order to contextualize our results within the broader HD142527 system architecture, Figure \ref{fig:disk_orb_det} superimposes (a) 50 randomly drawn orbits from the posterior distribution of orbit fits, (b) a single epoch VisAO detection of the H$\alpha$ point source, and (c) an H-band polarized intensity observation of the HD142527 outer disk. The image of the circumbinary disk was obtained by re-reducing GPI polarimetric data first published in \citet{Rodigas2014} using the updated GPI DRP described in \citet{DeRosa2020}; these data were then interpolated to VisAO's platescale to facilitate comparison with VisAO observations of the companion (shown for all epochs in Figure \ref{fig:SDIgallery}).

\par The ``mutual inclination" between the orbit of a binary star and its circumbinary disk is expected to inform the evolution of the circumbinary and circumstellar environments, and has wide-ranging implications for planet formation in binary systems \citep{Czekala2019}. While external binary companions may significantly truncate a circumstellar disk, perhaps suppressing planet occurrence \citep{Kraus2016}, circumbinary disks around tighter binary systems result in circumbinary planets \citep[e.g.][]{Kostov2020, Kostov2021}. The HD 142527 system is interesting to consider in this context because it is one of a handful of systems with direct imaging measurements of both inner and outer disk components as well as the binary orbit. 

The mutual inclination between the disk plane and binary orbit, $\theta$, is defined as 
\begin{eqnarray}
    \nonumber\cos{\theta}=\cos{i_{\mathrm{d}}}\cos{i_\star}+\sin{i_{\mathrm{d}}}\sin{i_\star}\cos{\left(\Omega_{\mathrm{d}}-\Omega_\star\right)},
\end{eqnarray}
\par where $\theta$ is the angle between the angular momentum vector of the binary orbit and the midplane of the disk \citep{Czekala2019}.  

\par We determine the posterior distribution of mutual inclinations between the binary orbit and both the inner and outer disk components (shown in Figure \ref{fig:mutualpost}) by drawing random samples from the posterior distribution of $i_\star$ and $\Omega_\star$ produced by \texttt{orbitize!}, and assuming a gaussian distribution for the disk orientation centered on the values most recently inferred from ALMA gas kinematic data for the outer disk \citep[$i_\mathrm{disk, outer}=38.21\pm1.38^\circ$, $\Omega_\mathrm{disk, outer}=162.72\pm1.38^\circ$,][]{Perez2015, Boehler2017, Bohn2021}, and VLTI/GRAVITY observations for the inner disk \citep[$i_\mathrm{disk, inner}=23.76\pm3.18^\circ$, $\Omega_\mathrm{disk, inner}=15.44\pm7.44^\circ$,][]{Bohn2021}. 

\par Regardless of the $180^\circ$ ambiguity in $\Omega_\star$, the current orbital solution for HD~142527~B yields large mutual inclinations throughout the system. We find $\theta_{\star-out} 89.84^{+2.30}_{-1.65}\mathrm{}^\circ$ or $ 158.82^{+2.76}_{-2.81}\mathrm{}^\circ$ (due to the mutual inclination's dependence on the bimodal $\Omega_\star$). These values are not consistent with those reported previously ($35\pm5 ^{\circ}$) by \citet{Czekala2019}, who adopted the $\Omega_\star$ solution from \citet{Claudi2019}. 
 
Our best fit to the mutual inclination of the binary with respect to the inner disk is $\theta_{in-\star}= 177.44^{+2.90}_{-2.80}\mathrm{}^\circ$ or $110.26^{+1.82}_{-1.50}\mathrm{}^\circ$.
\par Taken together, our mutual inclination fits present a system where each component of the system is dramatically misinclined with respect to the others.

\par Modeling the interaction of the binary companion and the circumbinary disk, \citet{Price2018} demonstrated that the morphological features of the circumbinary disk (wide cavity, asymmetric dust horseshoe, and spiral arms) can be qualitatively reproduced solely through interaction with the binary companion under orbits representative of those found in \citet{Lacour2016}. To best reproduce the observed morphology, \citet{Price2018} favor a narrower family of eccentric orbits ($\mathrm{e}=0.6-0.7$) with nearly perpendicular mutual inclinations with respect to the outer disk. 

\par Our work lends some support to this hypothesis, as one of the equal, symmetrically-distributed peaks in $\theta$ (resulting from the degeneracy in $\Omega_\star$) from our orbit fits centers on $\theta=89.84^{+2.30}_{-1.65}\mathrm{}^\circ$ suggestive of the near perpendicular configuration they describe. Our orbit fitting has not produced a distribution suggestive of such high eccentricities, however. 

\subsection{Photometry}
\par
\begin{figure*}
    \centering
    \includegraphics[width=0.8\textwidth]{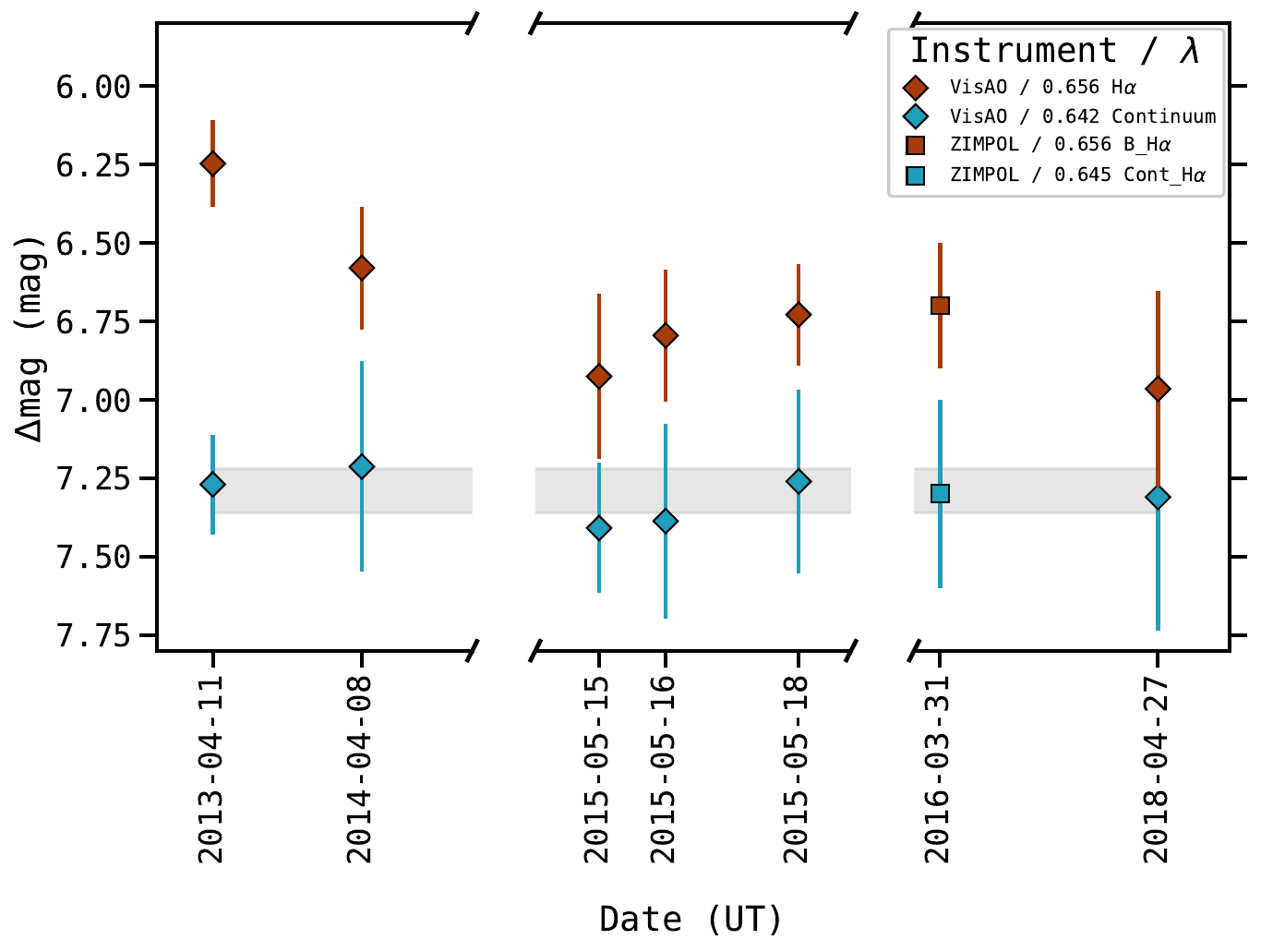}
    \caption{The visible-light contrast of HD~142527~B with respect to the continuum of HD~142527~A over time. Brown and green diamonds mark our H$\alpha$ and continuum contrast measurements, respectively. The grey shaded region represents the $1\sigma$ standard deviation of our measurements of continuum contrast, centered on the median. The red and blue squares mark the contrasts measured in nearly equivalent filters by \citet{Cugno2019} using VLT/SPHERE/ZIMPOL. The amount of H$\alpha$ excess appears to vary between many of the epochs, while the continuum contrast varies minimally and is consistent with uniformity within error bars.}
    \label{fig:142lightcurve}
\end{figure*}

 Figure \ref{fig:142lightcurve} plots the contrast of the companion \textit{relative to the stellar continuum} over time, derived as described in Section \ref{sec:results}. $\Delta\mathrm{mag}$ measurements from \citet{Cugno2019} are overplotted to demonstrate consistency with our measurements. Between 2013-04-11 and 2015-05-15, the H$\alpha$ contrast of the companion decreased by $0.68~\mathrm{mag}$. This is suggestive of a moderately variable accretion rate onto the companion on yearly timescales. Over the same baseline, the continuum did not vary within uncertainties. 
As there is not significant variability in the continuum contrast of the companion between epochs, this suggests that the extinction towards HD~142527~B is constant to $\pm0.2~\mathrm{mag}$. Accretion-driven variability in the stellar continuum flux at our cadence is similarly limited to at or below this value, though it likely occurs on shorter timescales \citep[e.g.][]{Stauffer2014}.

\par The \citet{Close2014} contrast measurements ($\Delta\mathrm{mag_{H\alpha}}=6.33\pm0.20~\mathrm{mag}$, $\Delta\mathrm{mag_{cont}}=7.50\pm0.25~\mathrm{mag}$), derived from the same dataset, agree with our measurements within error bars, and we note our slightly higher continuum brightness estimate is likely due to improved PSF subtraction and forward modeling, which allows us to better quantify flux lost to this process. While taken in marginally different filters, the \citet{Cugno2019} contrast measurements (shown with square symbols in Figure \ref{fig:142lightcurve}) agree well with our results; their continuum contrast measurement falls along our median continuum contrast and their H$\alpha$ contrast lies within the range of values we have measured, with similar uncertainties.
\par Our method for calculating the contrast of the companion with respect to the stellar continuum should should be insenstitive to stellar H$\alpha$ variability, leaving only stellar continuum variability as a potential contaminant. The star has a known periodicity of $\sim6$ days, with a peak-to-valley amplitude of $0.09~\mathrm{mag}$ in the R-band \citep{Claudi2019}, too small to account for the observed variation in the H$\alpha$ channel. If stellar continuum variability were contaminating the observed H$\alpha$ variation, we would expect to see it directly in our measurements of the companion's continuum contrast, and we do not observe such variation. We note that the companion's infrared continuum has been observed to vary on the order of 0.5mag \citep{Claudi2019} but we do not observe a similar variability at visible wavelengths.

\par Our data are therefore suggestive of variability in the H$\alpha$ emission of this directly imaged accreting companion, resulting in the observed time variability in the H$\alpha$ excess. This variability has implications for future direct imaging protoplanet surveys. If accretion onto less massive companions is similarly variable, detection limits will need to be interpreted with some caution, as companions will be more quiescent at certain times.

\subsection{Accretion Rate}
\par We estimate the mass accretion rate onto HD~142527~B using H$\alpha$ contrast measurements following a standard set of assumptions for accreting objects. First, we convert our contrast into a line luminosity measurement via \[\mathrm{L_{H\alpha}} = 4\pi D^2\times\mathrm{Z_{pt}}\times d\lambda\times10^{((\mathrm{mag_\star}+\Delta\mathrm{mag_B})/-2.5)},\] where $D$ is the distance to the system, $\mathrm{Z_{pt}}=2.339\times10^{-5}\mathrm{erg}/\mathrm{cm}^{2}/\mathrm{s}/\mu\mathrm{m}$ is the Vega zeropoint of the H$\alpha$ filter \citep{Males2014}, $d\lambda=0.006\mu\mathrm{m}$ is the width of the H$\alpha$ filter, and $\mathrm{mag_\star}=8.1-A_R$ is the R-band apparent magnitude of the central star \citep[8.1,][]{Cugno2019}, corrected for extinction \citep[$A_R=0.05$,][]{Fairlamb2015, Cugno2019}. 
\par We convert line luminosity to an estimate of the total accretion luminosity of the star using the \citet{Rigliaco2012} empirical relationship between $L_{H\alpha}$ and $L_{acc}$ derived for T-Tauri stars, namely: \[\log{L_\mathrm{acc}}=b+a\log{\left(L_{H\alpha}/L_\odot\right)},\] where $a=1.25\pm0.07$ and $b=2.27\pm0.23$. From the accretion luminosity, we derive the mass accretion rate via the standard relation: \[\dot{M} = \left(1-\frac{R_\star}{R_{in}}\right)^{-1} \frac{L_{acc}R_\star}{GM_\star}\sim1.25\frac{L_{acc}R_\star}{GM_\star},\] \citep{Gullbring1998} assuming $R_{in}\sim5R_\star$, as in \citet{Rigliaco2012}. We adopt the dynamical mass of the companion ($M_{\mathrm{B}} = 0.26M_\odot$) and best fit BHAC evolutionary model radius ($R_{\mathrm{B}} = 1.2R_\odot$) from \citet{Claudi2019} to compute our final accretion rate estimate from this equation.
\par Table \ref{tab:accretion} records the line luminosities and mass accretion rates for each epoch calculated following these assumptions. The peak H$\alpha$ excess, which occurs in the 2014-04-08 epoch, corresponds to a mass accretion rate estimate of $6\times10^{-10}~\mathrm{M_\odot}\mathrm{yr}^{-1}$.

\begin{deluxetable}{cccc}
\tablecaption{HD 142572B Accretion estimates\label{tab:accretion}}
\tablewidth{0pt}
\tablehead{
\colhead{Date} & \colhead{$L_{H\alpha}$} & \colhead{$\dot{M}$}\\
\colhead{MM/DD/YY} & \colhead{$L_\odot$} & \colhead{$\mathrm{M_\odot yr}^{-1}$}\\
}
\startdata
2013-04-11    & 2.12E-04 & 8.78E-10 \\
2014-04-08    & 1.56E-04 & 5.99E-10 \\
2015-05-15    & 1.14E-04 & 4.02E-10 \\
2015-05-16    & 1.28E-04 & 4.67E-10 \\
2015-05-18    & 1.36E-04 & 5.04E-10 \\
2018-04-27    & 1.10E-04 & 3.84E-10
\enddata
\end{deluxetable}

\par These mass accretion rate estimates differ from those calculated by \citet[$\mathrm{\dot{M}}=1-2\times10^{-10}~\mathrm{M_\odot}\mathrm{yr}^{-1}$]{Cugno2019}, who assumed a smaller radius \citep[$0.9\mathrm{R_\odot}$ given by][]{Lacour2016}. We assert that the larger radii inferred from the evolutionary model fit \citep{Claudi2019} is more appropriate given the purported age of the system, and this yields a slight increase in mass accretion rate estimate. The values still agree within an order of magnitude. 
\par Our derived values are consistent with the value of $5.9\times10^{-10}~\mathrm{M_\odot}\mathrm{yr}^{-1}$ reported by \citet{Close2014}; although they adopted a lower radius of $0.29\mathrm{R_\odot}$ and a mass of $0.25~\mathrm{M_\odot}$, they did not conduct the same accounting for the H$\alpha$ excess of the primary, which makes up the difference. Regardless of radius assumption, all derived accretion rates lie in a range consistent with observations of accretion rates onto low mass T-Tauri stars of similar masses and ages as HD~142527~B \citep[e.g. $10^{-9}-10^{-11}$,][]{Rigliaco2012}. 

\par Variability in mass accretion rates onto the primary stars of transitional disks has been observed previously. For example, the mass accretion rate of HD~142527~A changed by a factor of 7 over five years \citep{Mendigutia2014}. Variations in mass accretion rate on the order of factors of 2-10 on week-to-day long timescales have also been observed for accreting low mass stars more generally \citep[e.g.][]{Robinson2019}. We detect only marginally significant variability in the H$\alpha$ contrast of HD~142527~B given our conservative uncertainties. To our knowledge, however, this is the first detection of accretion variability in a secondary companion within a transitional disk gap.

\par HD~142527~B is likely surrounded by a circumsecondary disk \citep{Lacour2016} through which accreting material is processed. This circumsecondary disk is embedded within the cavity of the larger circumbinary disk, similar to the circumplanetary disk recently observed around PDS~70~c \citep{Benisty2021}. The non-detection results from \citet{Avenhaus2017} covered in \S\ref{sec:litrev} suggest that, if circumsecondary signal exists at the position of the companion, its contribution is much smaller than our photometric errors derived from Bayesian KLIP forward modeling. This motivates the need for a high-resolution ALMA search for circumsecondary material, as was recently done in the PDS~70 system \citep{Benisty2021}.

If the observed variability of HD~142527~B was due to the rotation of an accretion hotspot in and out of view, we would expect to see detectable continuum variability, similar to that observed in the NIR, but we do not. The observed variability could also be due to an accretion column rotating in and out of view. If that was the case, we might expect to observe large amplitude day-to-day variations between the 2015 observations, which we do not. The simplest explanation is that the accretion rate itself is variable over time, but more data are needed to verify this and to fully understand the degree and timescale of the variability of HD~142527~B.

\par Observations of accretion variability in planetary mass companions embedded within transition disk cavities have been difficult to  date. The H$\alpha$ lightcurve for PDS~70~b, for example, does not support large amplitude ($>30\%$) variability on month-long timescales \citep{Zhou2021}. Does this indicate that accretion onto objects orbiting within transition disk cavities is relatively stable when compared to accretion onto young stars? Recent modeling suggests that accretion onto a protoplanet from a circumplanetary disk may occur at a quasi-steady rate when averaged over week-long timescales, but should exhibit daily variability \citep{Takasao2021}. 

\par The difficulty in obtaining accurate photometry of accretion emission onto planetary mass companions limits our ability to detect variations in their H$\alpha$ excess. The next generation of instruments \citep[e.g. MagAO-X,][]{Males2020}, which can achieve higher Strehl ratios in the visible, may be able to observe such variability for a wider range of embedded companions using similar methods. Our result could indicate that accretion onto companions within cavities \textit{is} variable at a measurable level, at least for the highest mass companions. The caveat to this interpretation is that HD~142527~B is a stellar mass companion, whose accretion variability may be best understood in the domain of very-low-mass stellar accreting objects, rather than embedded planetary mass accretors.

\section{Conclusions}
\par In this paper, we present a five year monitoring campaign of the accreting companion HD~142527~B. We used the unique forward modeling capabilities of the Karhunen-Loeve Image Processing (KLIP) algorithm to achieve 1-2 mas precision on astrometric measurements taken over a 6 year time baseline, and validate an updated VisAO astrometric calibration solution (using the $\theta^1$ Ori B2-B3 binary, see Appendix \ref{sec:absastr}) by demonstrating good agreement between VisAO observations of HD~142527~B and the literature. We combine literature astrometry and multi-epoch MagAO/VisAO observations to fit the orbit of HD~142527~B using \texttt{orbitize!} and derive a posterior distribution of orbital elements. We verify that the companion is on an inclined ($\mathrm{i}=124.85\pm4.56^\circ$), eccentric ($\mathrm{e}=0.24\pm0.15$) orbit and is near periastron passage. We find that the HD142527 binary has a mutual inclination with respect to the outer disk of $\theta_{\star-out} = 89.84^{+2.30}_{-1.65}\mathrm{}^\circ$ or $158.82^{+2.76}_{-2.81}\mathrm{}^\circ$ (depending on the degenerate position of the ascending node, $\Omega$). We also find a dramatic mutual inclination with respect to the newly directly detected inner disk, $\theta_{in-\star} = 177.44^{+2.90}_{-2.80}\mathrm{}^\circ$ or $110.26^{+1.82}_{-1.50}\mathrm{}^\circ$. These newly derived inclinations could be used to guide hydrodynamical models of this system, in the context of disk warping, tearing, and precession in the presence of a disruptive companion. 
\par While HD~142527~B may be too tightly separated at the time of writing to be detected without interferometric techniques, astrometry of the companion following periastron passage ($\sim2021$) will be crucial in further constraining its orbit and therefore its mass. An improved mass estimate and updated characterization can yield, among other things, more precise accretion rate estimates. A single observation with the VLTI/GRAVITY interferometer, for instance, could track the companion's periastron passage, provide an improved mass constraint, and even place estimates (or upper limits) on the Br-$\gamma$ emission from HD~142527~B. With an improved mass constraint, the companion could be better compared to evolutionary models, yielding an improved age determination and a better understanding of the companion's formation. GRAVITY observations could also detect or place additional constraints on the size of the circumsecondary disk around HD~142527~B, as was done for the PDS~70 planets \citep{Wang2021}. High resolution ALMA observations could directly detect the presence of a circumsecondary disk around B, provide constraining astrometry, and explore the inner disk surrounding A. In the future, fiber-fed spectroscopy of the companion with an instrument such as KPIC may allow for the measurement of the radial velocity, absorption, and accretion signatures of the companion at high spectral resolution, which could help constrain its currently uncertain age and spectral type (M2.5-M7). 
\par Leveraging careful optimization of the \texttt{pyKLIP} algorithm, we achieved the most finely separated detection of a faint ($\Delta\mathrm{mag} > 6$) directly imaged companion using a non-coronagraphic, non-interferometric instrument to date.  We observe clear H$\alpha$ excess in all epochs of observation, corresponding to mass accretion rates similar to those observed in young, isolated M-dwarfs. We observe tentative signs of variability in the H$\alpha$ excess of the companion, suggestive of accretion variability. We estimate accretion rates for the HD~142527~B companion on the order of $4-9\times10^{-10}~\mathrm{M_\odot}\mathrm{yr}^{-1}$, assuming a radius based on evolutionary models.
\par Our results demonstrate that careful, long time-baseline observations from the current generation of high-contrast imaging instruments, combined with improvements in post-processing techniques, are able to place substantial constraints on both orbital motion and photometric variability, even for very tightly-separated directly imaged companions, provided a self-consistent data reduction and post-processing methodology. In the future, similar observations of systems such as PDS~70, LkCa 15, and AB Aur b (and other, newly discovered accreting protoplanets) with instruments such as MagAO-X will open new windows into the time variability of protoplanetary accretion and the process by which substellar companions form and evolve.

\vspace{1cm}
\section*{Acknowledgements}
\par We sincerely thank the anonymous reviewer for their thoughtful, rigorous, and supportive review that contributed immensely to the improvement of this paper. Special thanks to Gabriel-Dominique Marleau and Yuhiko Aoyama for their fruitful discussions of accretion physics. We thank Sarah Blunt for her tremendous help with our orbit fitting, Jason Wang for his \texttt{pyklip} expertise, as well as Connor Robinson for his encouragement. Special thanks to David Sing for his encouragement (and for letting WOB use his server to run last minute orbits). WOB would like to thank Yevaud, Kalessin, and Morgoth, as well as Benjamina, Martin, Luke, Emmett, Jack, and Kate Balmer.
\par WOB and KBF acknowledge funding from NSF-AST-2009816. WOB thanks the LSSTC Data Science Fellowship Program, which is funded by LSSTC, NSF Cybertraining Grant \#1829740, the Brinson Foundation, and the Moore Foundation; their participation in the program has benefited this work. KBF's work on this project was also supported by a NASA Sagan fellowship. LMC's work was supported by NASA Exoplanets Research Program (XRP) grants 80NSSC18K0441 and 80NSSC21K0397. KMM's work has been supported by the NASA XRP by cooperative agreement NNX16AD44G.
\par This paper includes data gathered with the 6.5 meter Magellan Telescopes located at Las Campanas Observatory, Chile. Some of the data presented herein were obtained at the W.M. Keck Observatory, which is operated as a scientific partnership among the California Institute of Technology, the University of California and the National Aeronautics and Space Administration. The Observatory was made possible by the generous financial support of the W.M. Keck Foundation.
\par This work has made use of data from the European Space Agency (ESA) mission {\it Gaia} (\url{https://www.cosmos.esa.int/gaia}), processed by the {\it Gaia} Data Processing and Analysis Consortium (DPAC, \url{https://www.cosmos.esa.int/web/gaia/dpac/consortium}). Funding for the DPAC has been provided by national institutions, in particular the institutions participating in the {\it Gaia} Multilateral Agreement.
\par WOB and KBF would like to acknowledge the land that the images used in this paper were observed from. Las Campanas Observatory, and the Magellan Clay Telescope, are built on Diaguita land. More on the Diaguita is available from the Museo Chileno de Arte Precolombino: \url{http://precolombino.cl/en/culturas-americanas/pueblos-originarios-de-chile/diaguita/}, and additional information from Diagutia activists can be found: \url{https://upndsalta.blogspot.com}. \textit{Astrobites} has written an informative article, ``Astronomical Observatories and Indigenous Communities in Chile," which can be found here: \url{https://astrobites.org/2019/09/10/astronomical-observatories-and-indigenous-communities-in-chile/}. The W. M. Keck Observatory, and the Keck II Telescope, are built on Native Hawaiian land. We are honored to be given the opportunity to conduct astronomy using data taken from this sacred place and would like to point to the informative \textit{Astrobites} articles (``Mauna Kea and Modern Astronomy," \url{https://astrobites.org/2018/11/09/mauna-kea-and-modern-astronomy/}, ``Maunakea, Western Astronomy, and Hawai`i," \url{https://astrobites.org/2019/08/02/maunakea-western-astronomy-and-hawaii/}. We would like to encourage our colleagues to seek out additional information about the on-going protests against additional construction on the mountain and the historical precedent for these protests.
\par The authors would like to acknowledge the land they have conducted research from during the course of this investigation. WOB and KBF would like to \href{https://www.fivecolleges.edu/academics/native-american-and-indigenous-studies#about-the-kwinitekw-valley}{acknowledge the Nonotuck land Amherst College occupies}, the Nonotuck ancestors, their descendants, and the neighboring Indigenous nations: the Nipmuc and the Wampanoag to the East, the Mohegan and Pequot to the South, the Mohican to the West, and the Abenaki to the North. WOB would also like to acknowledge the Piscataway community, their elders and ancestors, as well as their future generations. WOB acknowledges that \href{http://trujhu.org/index.php/about-us/land-acknowledgement/}{Johns Hopkins was founded upon the exclusions and erasures of many Indigenous peoples}, including those on whose land this institution is located.
\software{\texttt{astropy} \citep{astropy:2013, astropy:2018}, \texttt{emcee} \citep{emcee}, \texttt{photutils} \citep{photutils}, \texttt{ptemcee} \citep{ptemcee}, \texttt{pyklip} \citep{Wang2015}, \texttt{orbitize!} \citep{Blunt2020}}

\bibliography{balmer142}{}
\bibliographystyle{aasjournal}

\appendix
\vspace{-0.5cm}
\section{Astrometric Calibration of VisAO} \label{sec:absastr}
\par Given the precision ($\lesssim1~\mathrm{mas}$) afforded by BKA and other methods, the dominant source of error on the astrometry of directly imaged companions is often the absolute astrometric calibration of the instrument. In order to ensure our astrometric measurements of HD~142527~B were accurate when compared to literature astrometry, it was necessary to conduct an epoch to epoch calibration to determine the stability of the absolute astrometry of \ac{visao} over time.
\par The astrometric calibration of an instrument such as \ac{visao} is non-trivial due to the instrument's small FOV (8" x 8"), and because MagAO is a natural guide star adaptive optics system, which requires a bright target ($\mathrm{R_{mag}}<15$) within the field to make effective wavefront corrections. This significantly limits the number of viable calibration fields available to the observer.
\par Previous astrometric solutions for \ac{visao} \citep{Males2014, Close2013} have been calculated by cross-referencing \ac{magao}/\ac{visao} images with LBT/PISCES images of the dense Trapezium cluster \citep{Close2012}. The astrometric solution of PISCES was itself derived from cross-referencing against HST/ACS images of the cluster \citep{Ricci2008}. There is non-negligible orbital motion in these systems \citep{Close2013} that needs to be accounted for in images taken months to years apart. 
\par Additionally, \ac{visao} is removed and remounted between observing runs, which introduces epoch to epoch uncertainty in its calibration. Until now, \ac{visao} users have operated under the assumption that the astrometric calibration has remained stable between mountings since 2013, adopting the most recent published calibration of the instrument from \citet{Males2014}, who find a platescale of $7.8596 \pm 0.0019~\mathrm{mas}$ $\mathrm{pix}^{-1}$ and an offset from north of $0.59^\circ \pm \sim0.^\circ3$ clockwise based on Trapezium data collected in the Ys filter.
\par We measured the \ac{visao} astrometric solution by cross comparing \ac{visao} astrometry with an orbit fit to Keck/NIRC2 images\footnote{These data are available from the Keck Observatory Archive (KOA) at \url{koa.ipac.caltech.edu}.} of the $\theta^1$ Ori B2-B3 binary obtained between 2001 and Jan. 2020. The data were corrected for non-linearity, dark subtracted, flat fielded, and bad-pixel corrected before being corrected for geometric distortion, as described in \citet{DeRosa2020}. The NIRC2 astrometric solution is tied to precise HST/ACS observations of SiO masers in the Galactic Center \citep{Yelda2010, Service2016}.
\par
\begin{figure*}
    \centering
    \includegraphics[width=0.8\textwidth]{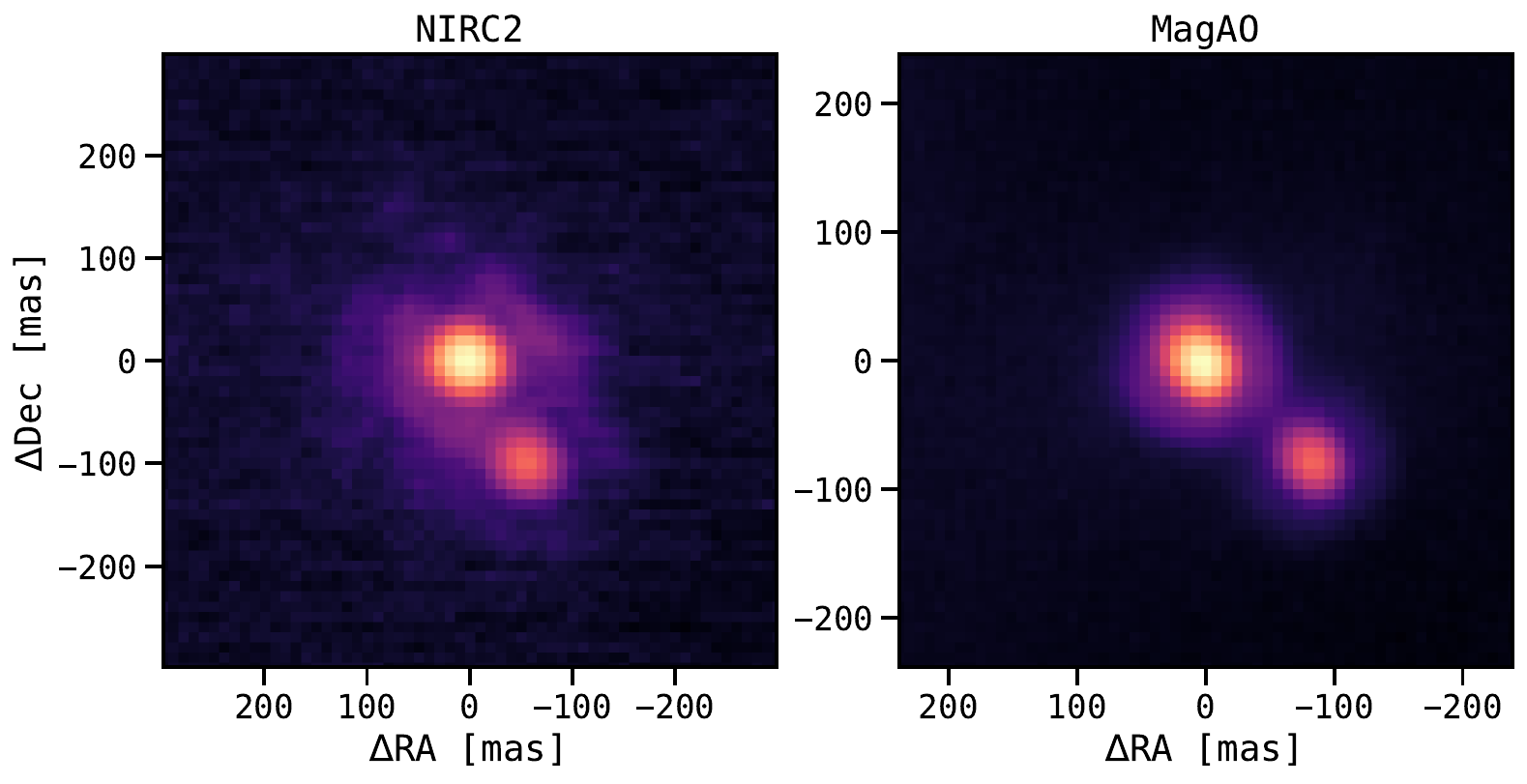}
    \caption{Representative images of $\theta^1$ Ori B2-B3 from NIRC2 (left, NB2.108 band) and MagAO (right, z' band). All relative astrometry herein is conducted with respect to $\theta^1$ Ori B2, which is centered at (0,0).}
    \label{fig:tet1ori_imgs}
\end{figure*}

 Our calibration method follows \citet{DeRosa2020}, who conducted a similar calibration of the Gemini Planet Imager (GPI) using the same Keck/NIRC2 images. We measure the position of the $\theta^1$ Ori B2-B3 binary across more than 20 years of NIRC2 imagery using \texttt{photutils} PSF photometry \citep{photutils}. We used the PSF of $\theta^1$ Ori B1 to conduct PSF photometry, fitting both components of the binary jointly with \texttt{orbitize!}, which allows us to determine the expected position of the binary at arbitrary epochs and accounts for non-linear orbital motion over time. Our observations of HD 142527 and our VisAO calibration images of the $\theta^1$ Ori B system fall within the NIRC2 baseline of observations, meaning that each comparison is made within a well-constrained orbital fit. Figure \ref{fig:tet1ori_imgs} shows two representative images of $\theta^1$ Ori B2-B3 from NIRC2 and MagAO respectively.
\par We initialized an \texttt{oribitze!} MCMC with 50 walkers, which each took 100,000 steps after 100,000 ``burn-in" steps were discarded, for a total of 5 million accepted fit orbits. We set the initial total system mass to be $5.5\pm0.5M_\odot$ \citep{Close2013} and the initial system parallax to $2.655\pm0.042$ \citep{GaiaCollaboration2021}. We then determined the platescale and North Angle offset in each dataset by comparing the separation and Position Angle of the B2-B3 binary in VisAO images of the system to the expected separation and Position Angle of the NIRC2 orbital fit. The VisAO images of the $\theta^1$ Ori B system were reduced and registered following the methods described in Section \ref{sec:observations}, and the positions measured using \texttt{photutils} PSF photometry.
\par We determined the platescale and North Angle offset for each dataset, then took the average of each dataset weighted by the measurement uncertainty. We observe no significant variations in platescale or North Angle offset over time or between filters, and conclude that both have remained stable across instrument mountings. Our measurements of each therefore represent observations of a constant over time, and we take the weighted average of measurements from individual epochs to be the most precise estimate of the true platescale and North Angle for \ac{visao}.
\par We find the updated \ac{visao} platescale to be $7.95\pm0.010 ~\mathrm{mas}$ $\mathrm{pix}^{-1}$ and the North Angle offset to be $0.497\pm0.192^\circ$ counterclockwise. Our updated calibration yields updated errors, and we find (similar to \citeauthor{DeRosa2020}) slightly larger errors on the platescale than previous calibrations, and smaller errors on the North Angle offset. It would appear that by measuring the platescale in only one filter with a limited number of observations, previous calibrations underestimated the platescale error, and that a longer baseline of observations allows improved precision in determination of the North Angle offset. 
\par
\begin{figure*}
    \centering
    \includegraphics[width=\textwidth]{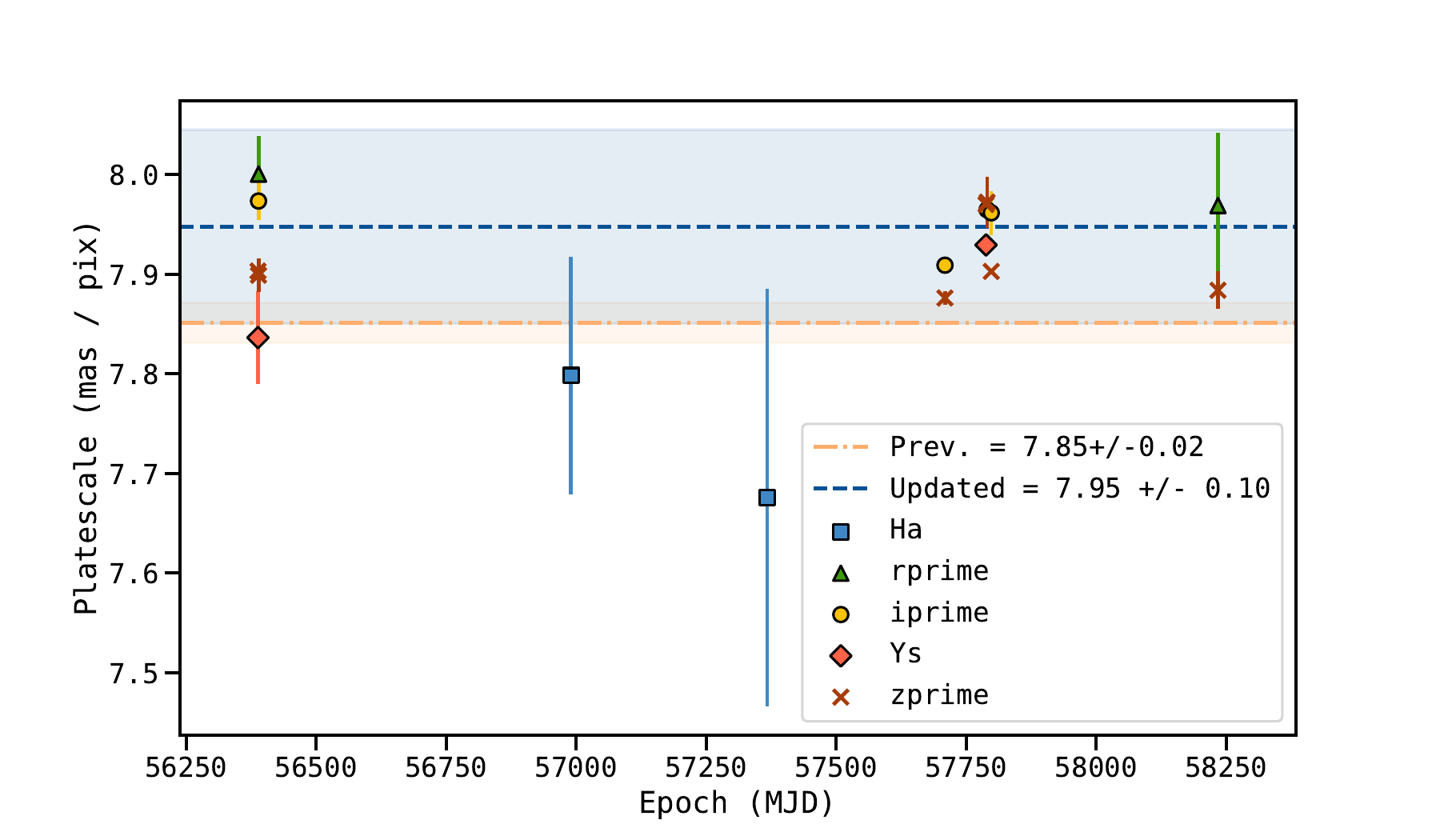}
    \caption{VisAO platescale as measured in various VisAO filters over time. The orange shaded region, centered on the horizontal orange dash-dot line, represents the previous astrometric solution of \citet{Males2014} using data taken in the Ys filter in 2014. Note the agreement between our measurement of the same Ys data (red diamond) and their value. We note that the H$\alpha$ observations suffered from poor observing conditions and an unfavorable observing strategy for astrometry which resulted in very low signal-to-noise on the B2-B3 pair. No obvious trend in platescale with wavelength or time is present, and therefore we adopt the weighted average and standard deviation on the weighted average (represented by the blue dashed line and shaded region, respectively) as the updated platescale and platescale error for the instrument.}
    \label{fig:platescale}
\end{figure*}

\begin{figure*}
    \centering
    \includegraphics[width=\textwidth]{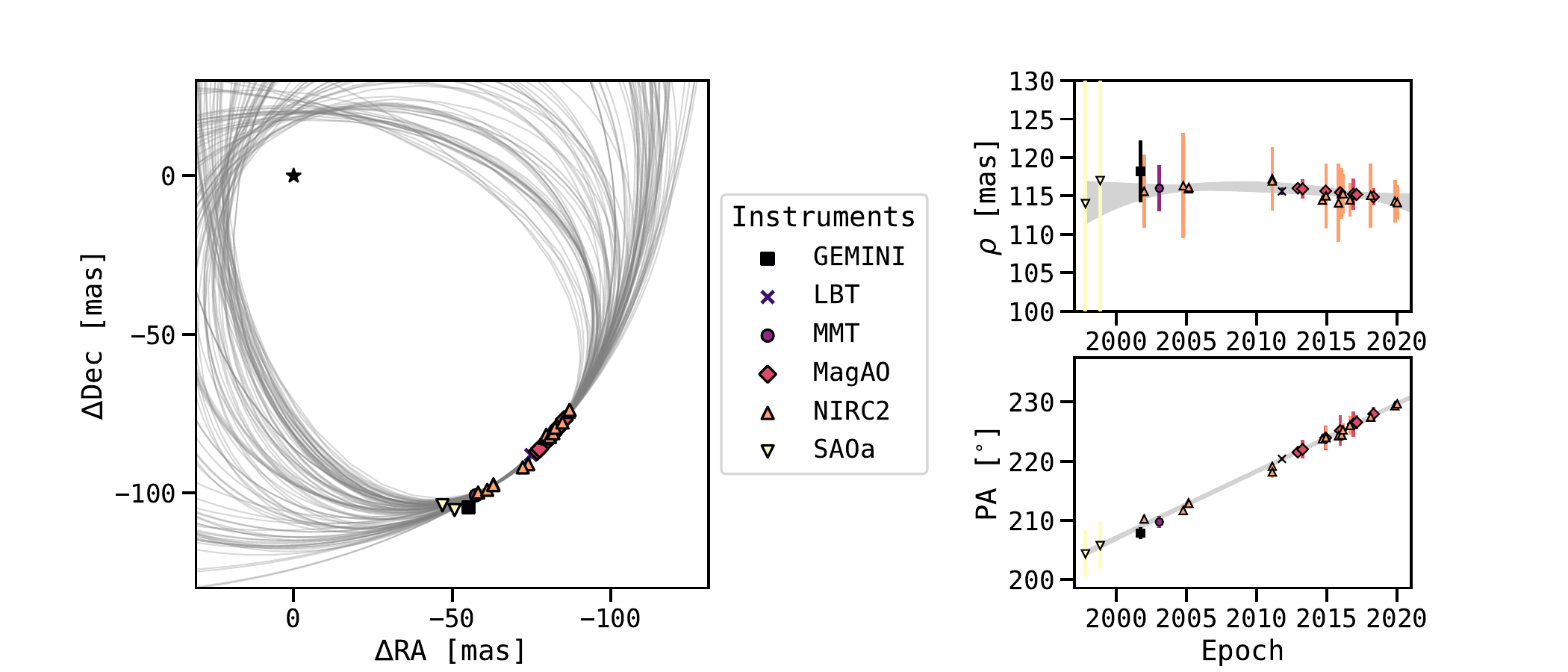}
    \caption{500 orbital fits to the NIRC2 astrometry of $\theta^1$ Ori B2-B3 randomly drawn from the posterior distribution. Astrometry compiled in Table \ref{tab:b2b3} is overplotted. Only the NIRC2 astrometry (orange triangles) was used to fit the orbit of the binary, and the corrected VisAO astrometry (red diamonds) falls along the fit orbits by design.}
    \label{fig:b2b3orbit}
\end{figure*}

 Figure \ref{fig:platescale} plots measurements of the VisAO platescale over time and across multiple filters. Figure \ref{fig:b2b3orbit} shows the orbital fit to the NIRC2 observations of $\theta^1$ Ori B2-B3 and updated VisAO astrometry of the binary pair.

\par As a byproduct of this astrometric calibration, we have obtained astrometric measurements of the $\theta^1$ Ori B2-B3 over nearly 20 years, as well as an updated, well-defined orbital solution for the binary. We record the astrometry from both NIRC2 and VisAO with other literature astrometry in Table \ref{tab:b2b3}. We record the orbital elements as drawn from the posterior distribution of fits to the NIRC2 astrometry in Table \ref{tab:b2b3elems}.

\begin{deluxetable}{ccccc}
\tablecaption{$\theta^1$ Ori B2-B3 astrometry \label{tab:b2b3}}
\tablewidth{0pt}
\tablehead{
\colhead{Epoch} & \colhead{Separation} & \colhead{Position Angle} & \colhead{Instrument} & \colhead{PI} \\
\colhead{(MJD)} & \colhead{(mas)} & \colhead{($^\circ$)} & \colhead{} & \colhead{}
}
\startdata
50735       & $114.000\pm0.050$   & $204.300\pm4.000$  & SAOa       & Weigelt et al. 1999 \\
51120       & $117.000\pm0.005$   & $205.700\pm4.000$  & SAOa       & Weigelt et al. 1999 \\
52171       & $118.200\pm0.004$   & $207.800\pm1.000$  & GEMINI     & Close et al. 2003   \\
52263       & $115.600\pm4.766$   & $210.212\pm0.595$  & NIRC2      & This work           \\
52659       & $116.000\pm0.003$   & $209.700\pm1.000$  & MMT        & Close et al. 2003   \\
53281       & $116.329\pm6.831$   & $211.649\pm0.768$  & NIRC2      & This work           \\
53417       & $115.919\pm0.444$   & $212.922\pm0.165$  & NIRC2      & This work           \\
53426       & $116.109\pm0.112$   & $212.840\pm0.030$  & NIRC2      & This work           \\
55850       & $115.600\pm0.001$   & $220.390\pm0.300$  & LBT        & Close et al. 2012   \\
55598       & $117.249\pm4.133$   & $219.128\pm0.543$  & NIRC2      & This work           \\
55598       & $116.936\pm1.880$   & $218.154\pm0.943$  & NIRC2      & This work           \\
56265       & $116.000\pm0.000$   & $221.500\pm0.300$  & MagAO      & Close et al. 2013   \\
56388       & $115.914\pm1.223$   & $221.989\pm1.584$  & MagAO      & This work           \\
56389       & $115.914\pm1.015$   & $221.992\pm1.047$  & MagAO      & This work           \\
56990       & $115.648\pm0.550$   & $223.931\pm2.079$  & MagAO      & This work           \\
56903       & $114.453\pm0.557$   & $223.776\pm0.860$  & NIRC2      & This work           \\
56997       & $115.012\pm4.235$   & $223.980\pm1.871$  & NIRC2      & This work           \\
57322       & $114.090\pm5.086$   & $224.290\pm0.694$  & NIRC2      & This work           \\
57367       & $115.447\pm1.301$   & $225.152\pm2.603$  & MagAO      & This work           \\
57405       & $115.318\pm3.311$   & $224.424\pm0.662$  & NIRC2      & This work           \\
57405       & $115.318\pm3.311$   & $224.424\pm0.662$  & NIRC2      & This work           \\
57439       & $115.296\pm2.548$   & $225.253\pm1.120$  & NIRC2      & This work           \\
57620       & $114.498\pm2.178$   & $226.040\pm1.549$  & NIRC2      & This work           \\
57709       & $115.239\pm2.008$   & $226.264\pm2.108$  & MagAO      & This work           \\
57788       & $115.188\pm0.116$   & $226.521\pm0.498$  & MagAO      & This work           \\
57790       & $115.187\pm0.117$   & $226.527\pm0.526$  & MagAO      & This work           \\
57798       & $115.182\pm0.673$   & $226.554\pm1.218$  & MagAO      & This work           \\
58162       & $115.067\pm4.193$   & $227.458\pm0.558$  & NIRC2      & This work           \\
58234       & $114.875\pm1.121$   & $227.979\pm1.218$  & MagAO      & This work           \\
58790       & $114.316\pm2.814$   & $229.360\pm0.561$  & NIRC2      & This work           \\
58852       & $114.130\pm2.234$   & $229.690\pm0.003$  & NIRC2      & This work          
\enddata
\tablecomments{SAOa speckle interferometry from \citet{Weigelt1999}, and GEMINI, MMT, and LBT direct imaging from \citet{Close2003, Close2012, Close2013}}. Note that the MagAO measurements presented here are not independent measurements, as they were tied to the orbit fit to NIRC2 observations in order to determine the platescale and North Angle offset.
\end{deluxetable}

\begin{deluxetable}{cc}
\tablecaption{$\theta^1$ Ori B2-B3 orbital elements\label{tab:b2b3elems}}
\tablewidth{0pt}
\tablehead{
\tablehead{\colhead{Element} & \colhead{Value}}
}
\startdata
a (au)                                             & $42.23_{-1.38}^{+2.83}$ \\
e                                              & $0.72_{0.02}^{+0.02}$ \\
i $(^\circ)$                                   & $53.64_{-0.90}^{+0.98}$ \\
$\omega (^\circ)$                              & $91.09\pm2.26, 271.05\pm2.31$  \\
$\Omega (^\circ)$                              & $114.18\pm13.00, 293.07\pm13.36$ \\
Periastron (yr)                                & $2059.17_{-4.07}^{+5.11}$ \\
Parallax (mas)                                 & $2.65_{-0.04}^{+0.04}$ \\
$\mathrm{M_{total}} (M_\odot)$                 & $5.44_{-0.46}^{+0.48}$ \\
\enddata
\end{deluxetable}

\vspace{-1.3cm}
\section{Forward Model choice and VisAO ghost calibration} \label{sec:ghost}
\par
\begin{figure*}
    \centering
    \includegraphics[width=0.6\textwidth]{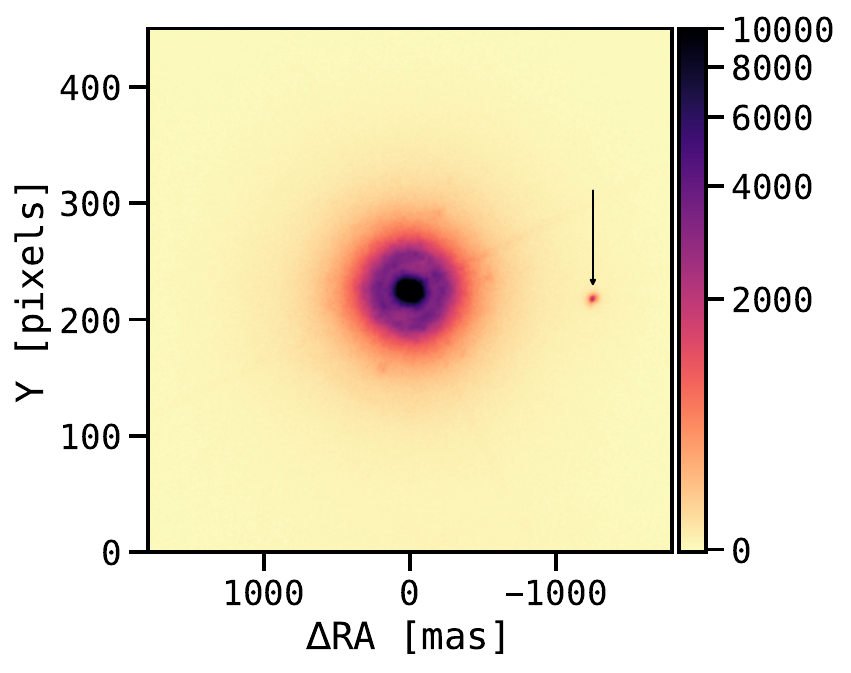}
    \caption{A saturated image of HD 142527. The position of the instrumental ghost is marked with a black arrow. The Y axis is in pixels, and the X axis is labeled in milliarcseconds to illustrate the spatial extent of the 451 pixel crop and the position of the ghost.}
    \label{fig:ghostex}
\end{figure*}

 The VisAO CCD saturates at $\sim16,000$ counts, and because of its small FOV, there are no other stars within the field that can be used as a PSF to forward model. There are no artificial satellite spots injected as astrometric or photometric calibrators, but there is an instrumental `ghost' PSF that appears to the right of the natural guide star in each image (see Figure \ref{fig:ghostex}). We investigated the stability of the ghost and the scaling relationships between the ghost and central PSF using 10 unsaturated datasets taken as part of the GAPlanetS survey \citep[and Follette et al. 2022, in prep.]{Follette2017}. 
\par We find that peak of the ghost varies consistently with the peak of the central PSF. We adopt the empirical scaling relationships $\mathrm{F_{c, H\alpha}}=179.68\pm4.59\times\mathrm{F_{g, H\alpha}}$ and $\mathrm{F_{c, cont.}}=196.31\pm3.56\times\mathrm{F_{g, cont.}}$ However, we find that when fit with a Moffat distribution, the FWHM of the ghost is on average 7\% larger than that of the central PSF. We therefore adopt the relationship $\mathrm{FWHM_c} = 0.93\times\mathrm{FWHM_{ghost}}$ in determining the FWHM of saturated images.
\par This ghost calibration was conducted in part because we had initially used the instrumental ghost as a forward model PSF, and found it to be a poor fit. We then investigated the best choice of forward model for saturated data by comparing sum-of-squares residuals for the ghost itself, a Moffat distribution fit to the ghost, a Gaussian distribution fit to the ghost, and both distributions (Moffat and Gaussian) with reduced FWHM equal to $0.93\times\mathrm{FWHM_{ghost}}$. For each PSF, we conducted the same procedure as in Section \ref{sec:results}, forward modeling the PSF through KLIP and fitting it to the known companion, resulting in a posterior distribution of fits and a residual map. We found that for the majority of our images of HD~142527~B, Gaussian PSFs yielded the smallest residuals. We assume that, scaling by the above relationships, the counts under the Gaussian are equivalent to those under the unsaturated central PSF. This then enables us to conduct photometry using BKA.

\section{Forward Model Fits} \label{sec:BKAfits}

\par
\begin{figure*}
    \centering
    \includegraphics[height=\textheight]{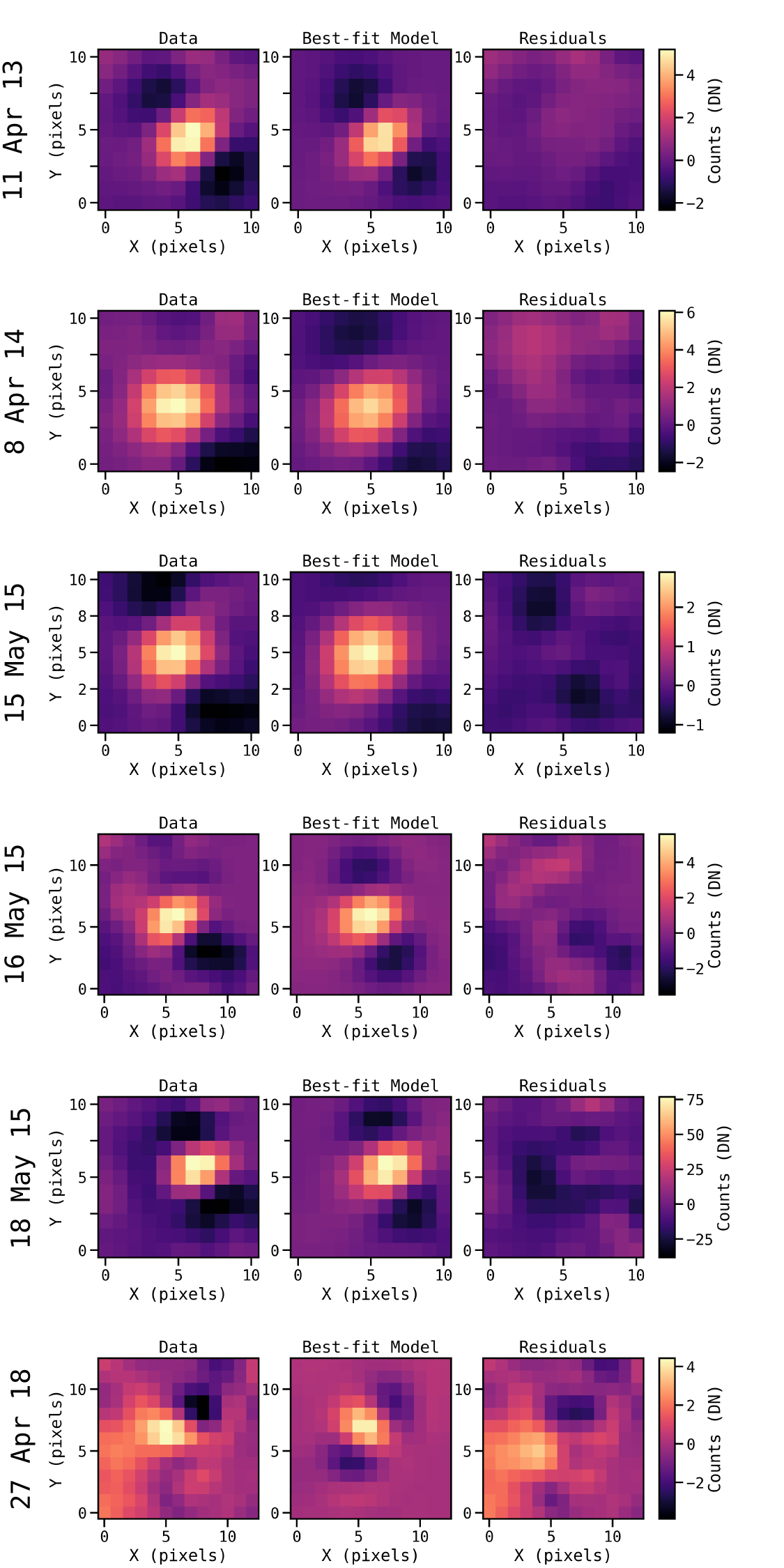}
    \caption{A gallery of BKA forward model best fits to HD~142527~B in the H$\alpha$ filter, in chronological order from top to bottom. Data (left) is fit by BKA, yielding a best fit forward model (center), and their difference (right).}
    \label{fig:haBKAgallery}
\end{figure*}

\begin{figure*}
    \centering
    \includegraphics[height=\textheight]{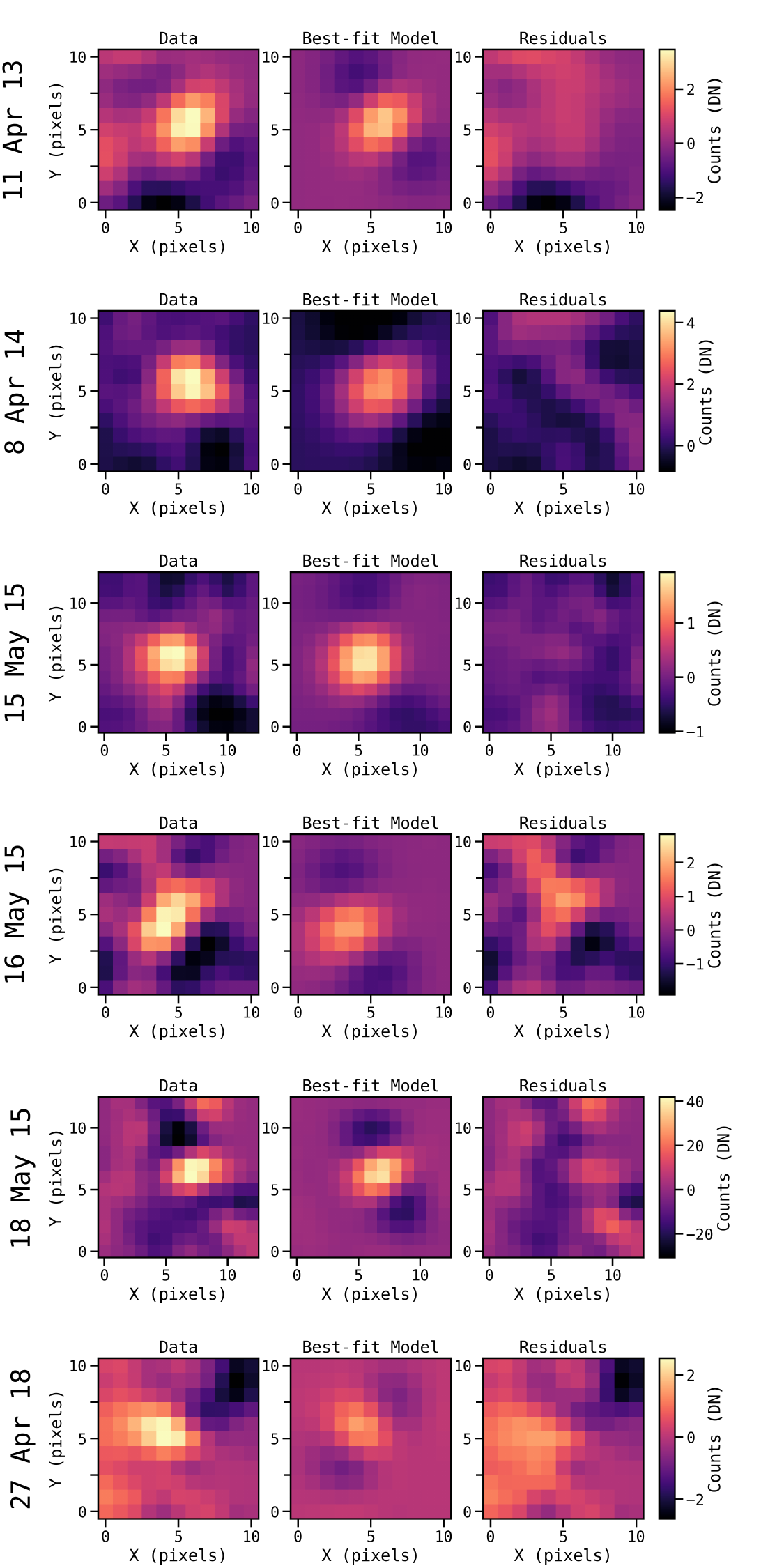}
    \caption{A gallery of BKA forward model best fits to HD~142527~B in the continuum filter, in chronological order from top to bottom. Data (left) is fit by BKA, yielding a best fit forward model (center), and their difference (right).}
    \label{fig:contBKAgallery}
\end{figure*}

\begin{figure*}
    \centering
    \includegraphics[width=\textwidth]{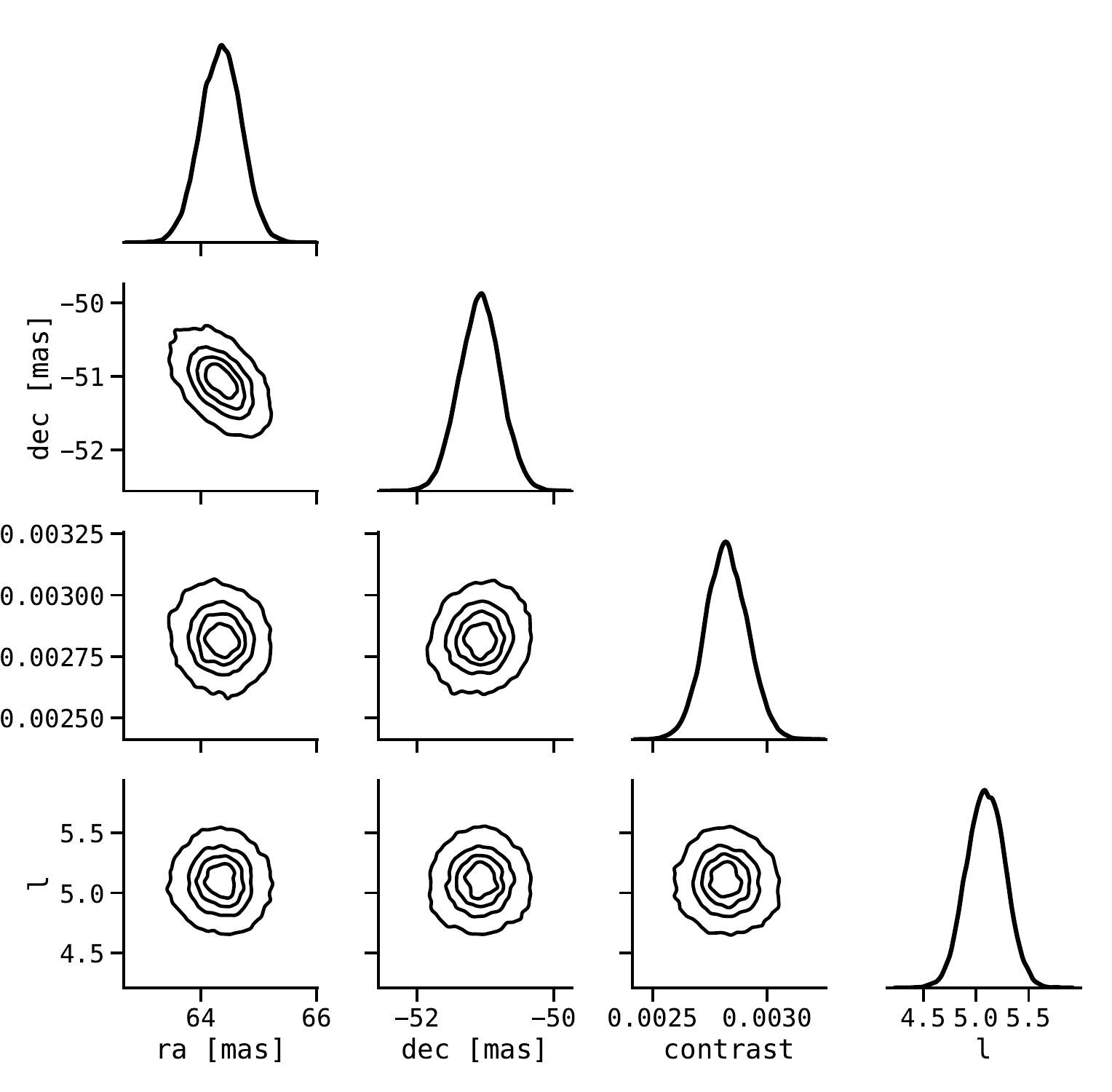}
    \caption{A corner plot illustrating the posterior distribution of BKA forward model fits to the 2013-04-11 H$\alpha$ data, which is representative of all other epochs. Note that all marginal parameters are normally distributed, and the only correlation is a slight linear correlation between X and Y position.}
    \label{fig:bkacornerex}
\end{figure*}

 Optimized post-KLIP images for each detection epoch, best fit BKA models, and the residuals between them are shown for HD~142527~B in the H$\alpha$ and continuum filters in Figures \ref{fig:haBKAgallery} and \ref{fig:contBKAgallery}, respectively. The marginal posterior distributions for the BKA fits, an example of which is shown in Figure \ref{fig:bkacornerex}, are distributed normally for all epochs. The most common correlation is a slight linear correlation between X and Y position (as seen in Figure \ref{fig:bkacornerex}).

\section{Orbit fit posteriors} \label{sec:posteriors}
\par
\begin{figure*}
    \centering
    \includegraphics[width=\textwidth]{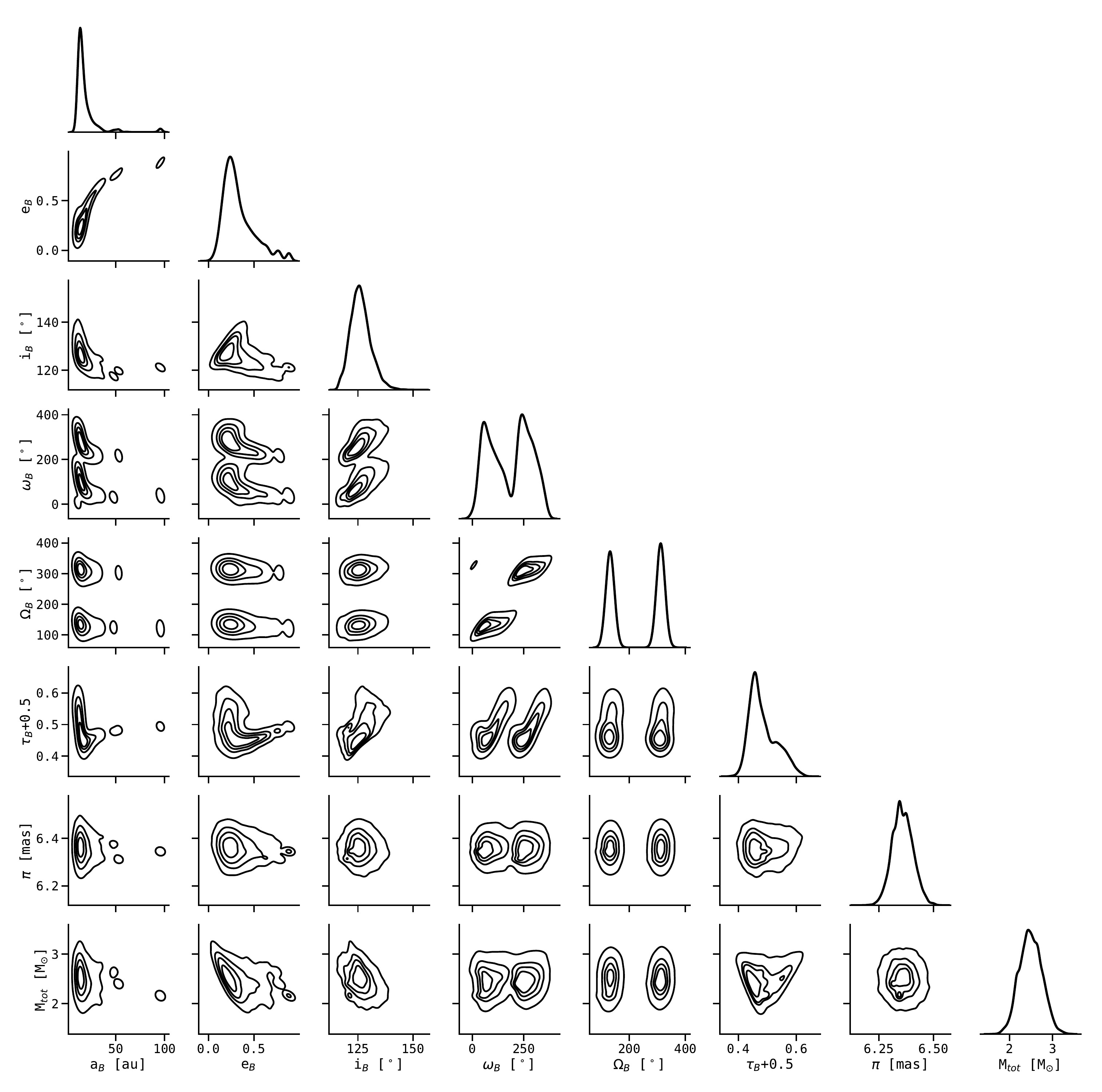}
    \caption{Posterior distribution of orbital elements fit to the astrometry of HD~142527~B. $\omega$ and $\Omega$ show bimodal distributions with peaks spaced $180^\circ$ apart; this a known degeneracy in visual orbits with a lack of RV constraints.}
    \label{fig:142post}
\end{figure*}

\begin{figure*}
    \centering
    \includegraphics[width=\textwidth]{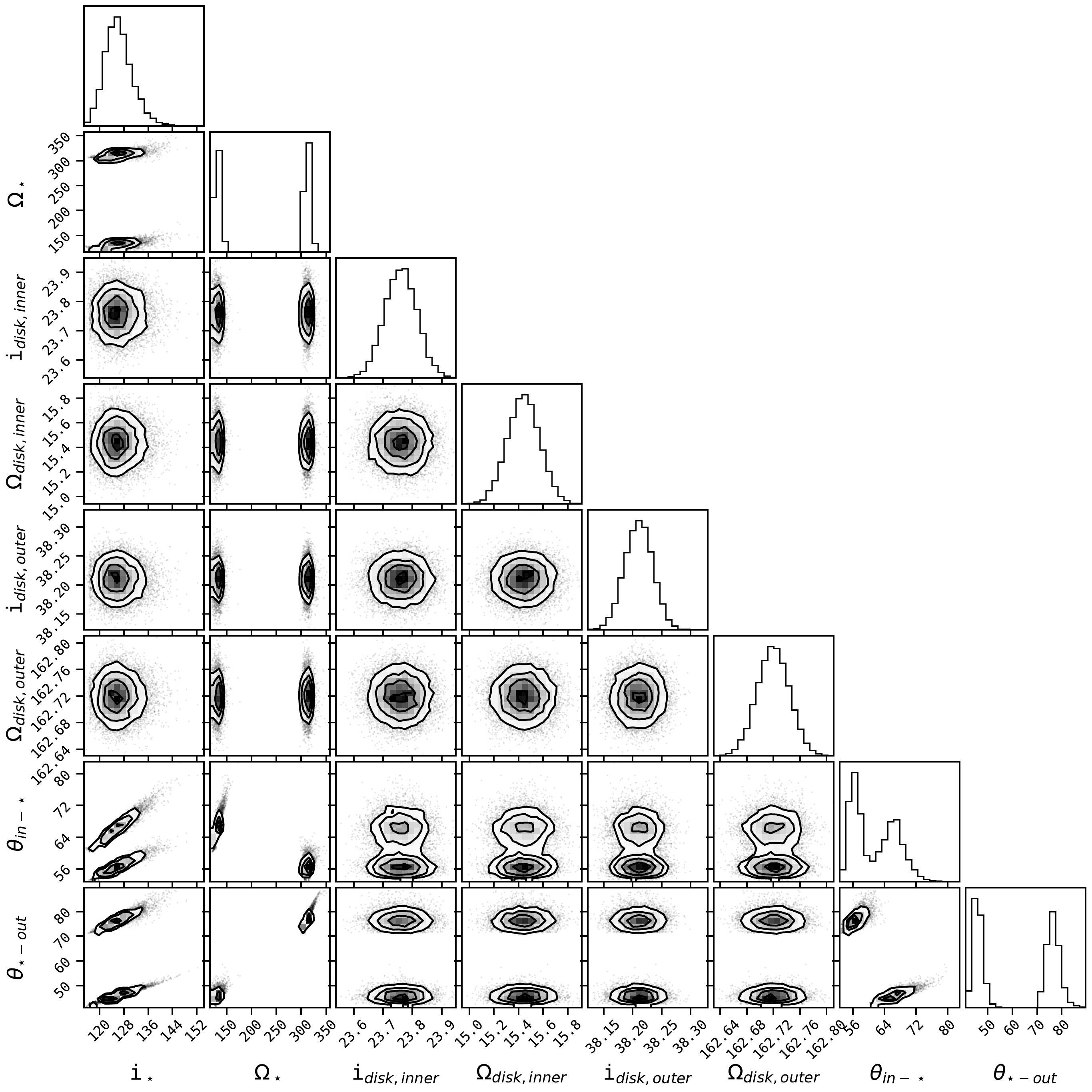}
    \caption{Posterior distribution of fit ($\mathrm{i_\star}, \Omega_\star$) and randomly sampled ($\mathrm{i_{disk}}, \Omega\mathrm{_{disk}}$) for both inner and outer disks yields a posterior distribution of $\theta$, the mutual inclination angle between the binary orbit and the disk component. As in \citet{Czekala2019}, the mutual inclination of the outer disk for this system is multimodal, but dramatically misaligned ($\theta\gg3^\circ$) no matter choice of $\Omega_\star$. Interestingly, one family of $\theta_{\star - \mathrm{out}}$ is nearly perpendicular, similar to the configuration described in \citet{Price2018}.}
    \label{fig:mutualpost}
\end{figure*}

\begin{figure*}
    \centering
    \includegraphics[width=\textwidth]{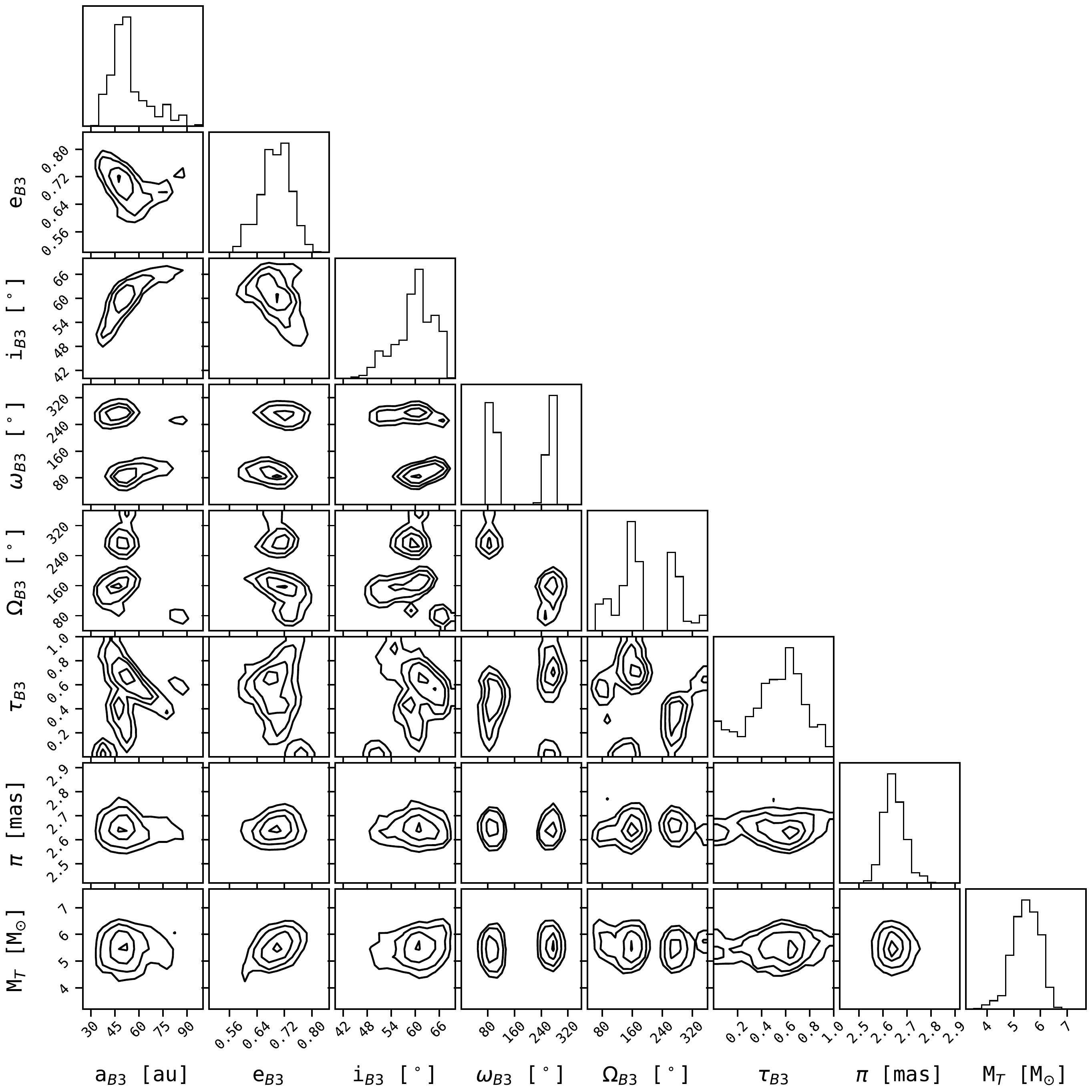}
    \caption{Posterior distribution of orbital elements fit to NIRC2 astrometry of $\theta^1$ Ori B2-B3. The fit displays a posterior distribution typical of short-arc visual orbit fits, with the same bimodal $\omega$ and $\Omega$ noted previously, normally distributed $\pi$ and $\mathrm{M_{tot}}$, correlated semi-major axis, eccentricity, and inclinations that are otherwise relatively constrained. This orbit is so satisfyingly typical, I have petitioned for it act as the unofficial mascot of \texttt{orbitize!} but I have not yet been humored thusly.}
    \label{fig:b2b3post}
\end{figure*}

 This appendix details posterior distributions of orbital elements for our astrometric fits, computed using \texttt{orbitize!}. Figure \ref{fig:142post} illustrates the posterior distribution of orbital elements for the HD~142527~AB binary. Figure \ref{fig:mutualpost} plots the mutual inclination parameters $\mathrm{i_\star}, \Omega_\star$ from Figure \ref{fig:142post}, computed as described in Section \ref{sec:analysis}, along with $\mathrm{i_d}, \Omega\mathrm{_d}$ drawn from gaussian distributions specified by the disk parameters fit by \citet{Bohn2021}, and the resultant distribution of mutual inclinations $\theta$. Figure \ref{fig:b2b3post} illustrates the posterior distribution for the orbit of the $\theta^1$ Ori B2-B3 derived from Keck NIRC2 observtions, as described in Appendix \ref{sec:absastr}. 

\end{document}